\documentclass[]{raa} %referee          

\usepackage{graphicx,times}
\usepackage{natbib}
\usepackage{amssymb,amsmath}
\usepackage[figuresright]{rotating}
\usepackage{pdflscape}
\usepackage{siunitx}
\usepackage{longtable}
\bibpunct{(}{)}{;}{a}{}{,}

\newcommand{\soft}{\texttt}
\newcommand{\HeTu}{\soft{HeTu}}
\newcommand{\astroquery}{\soft{Astroquery}}

\usepackage[pagebackref=true]{hyperref}
\hypersetup{
	colorlinks=true,
	linkcolor=blue,     
	urlcolor=blue,
	citecolor=blue,
}

\begin{document}
   \title{A Machine Learning made Catalog of FR-II Radio Galaxies from the FIRST Survey}

 \volnopage{ {\bf 20XX} Vol.\ {\bf X} No. {\bf XX}, 000--000}
   \setcounter{page}{1}

   \author{Bao-Qiang Lao\inst{1}, Xiao-Long Yang\inst{2}, Sumit Jaiswal\inst{2}, Prashanth Mohan\inst{2}, Xiao-Hui Sun\inst{1}, Sheng-Li Qin\inst{1}, Ru-Shuang Zhao\inst{3}
   }

   \institute{ School of Physics and Astronomy, Yunnan University,
             Kunming 650091, China; {\it lbq19881213@gmail.com}\\
        \and
             Shanghai Astronomical Observatory, Chinese Academy of Sciences,
            Shanghai 200030, China; {\it yangxl@shao.ac.cn}\\
 	\and
             School of Physics and Electronic Science, Guizhou Normal University, Guiyang 550001, China \\
% %	  Center for Astrophysics, University of Science and Technology of China, Hefei 230026, China\\
% Key Laboratory for Research in Galaxies and Cosmology, The University of Science
% and Technology of China, Chinese Academy of Sciences, Hefei, Anhui, 230026, China\\
% \and 
% Polar Research Institute of China,
% Jinqiao Rd. 451, Shanghai, 200136, China\\
\vs \no
   {\small Received 20XX Month Day; accepted 20XX Month Day}
}

\abstract{We present an independent catalog (FRIIRGcat) of 45,241 Fanaroff-Riley Type II (FR-II) radio galaxies compiled from the Very Large Array Faint Images of the Radio Sky at Twenty-centimeters (FIRST) survey and employed the deep learning method. Among them, optical and/or infrared counterparts are identified for 41,425 FR-IIs. This catalog spans luminosities $2.63\times10^{22}\leq L_{\rm rad}\leq6.76\times10^{29}\,{\rm W}\,{\rm Hz}^{-1}$ and redshifts up to $z=5.01$. The spectroscopic classification indicates that there are 1431 low-excitation radio galaxies and 260 high-excitation radio galaxies. Among the spectroscopically identified sources, black hole masses are estimated for 4837 FR-IIs, which are in $10^{7.5}\lesssim M_{\rm BH}\lesssim 10^{9.5}$ $M_{\odot}$. Interestingly, this catalog reveals a couple of giant radio galaxies (GRGs), which are already in the existing GRG catalog, confirming the efficiency of this FR-II catalog. Furthermore, 284 new GRGs are unveiled in this new FR-II sample; they have the largest projected sizes ranging from 701 to 1209 kpc and are located at redshifts $0.31<z<2.42$. Finally, we explore the distribution of the jet position angle and it shows that the faint Images of the FIRST images are significantly affected by the systematic effect (the observing beams). The method presented in this work is expected to be applicable to the radio sky surveys that are currently being conducted because they have finely refined telescope arrays. On the other hand, we are expecting that further new methods will be dedicated to solving this problem.
}

   \authorrunning{B.-Q. Lao et al. }            %author_head in even pages
   \titlerunning{A Machine Learning made Catalog of FR\,II Radio Galaxies from the FIRST Survey}  % title_head in odd pages
   \maketitle

%________________________________________________ sections below
% 
\section{Introduction}           %% first-level sections will be auto-capitalized
\label{sect:intro}
Supermassive black holes (SMBHs, black hole masses range $10^6 - 10^{10}\,M_\odot$) are generally identified at the centers of massive galaxies with bulges (\citealt{1999PNAS...96.4749F,2017A&ARv..25....2P}). Accretion and ejection are the primary processes in which the central SMBHs interplay with their host galaxies (\citealt{2017NatAs...1E.165H,2021RAA....21..212Z}), which form the structure of accretion discs and jets/outflows simultaneously. Accreting SMBHs, a.k.a. Active Galactic Nucleus (AGNs), are highly energetic astronomical objects at all electromagnetic wavelengths from radio to Gamma-ray. Especially, the synchrotron emissions from AGN jets dominate the radio band. According to the geometrical conditions of jets, radio-loud AGNs can be divided into exactly different types, i.e. Blazars and radio galaxies. 

Radio galaxies can present compact or extended radio morphologies (\citealt{2017JPhCS.869a2078M}), with the extended galaxies traditionally classified into two main classes, namely Fanaroff-Riley (FR) Type I (FR-I) and Type II (FR-II) (\citealt{fanaroff1974morphology}). Based on the FR definition, those displaying a bright core or an ``edge-darkened" characteristic along with diffuse lobes are classified as FR-Is, and those where the features away from the core are dominated by ``edge-brightened" characteristics are referred to as FR-IIs. This morphological classification scheme was found to be correlated with their radio luminosity. Fanaroff and Riley observed that almost all the sources in their samples with a radio luminosity smaller than $2\times10^{25}\,{\rm W}\,{\rm Hz}^{-1}$ were classified as FR-Is, while the brighter sources were predominantly identified as FR-IIs.

To unveil the optical features of radio galaxies, an alternative classification scheme was proposed by \cite{1994ASPC...54..201L} based on the intensity of optical emission lines. This scheme utilizes the relative intensity of high- and low-excitation lines, such as the [O III] emission line, in optical spectra and classifies sources into high-excitation radio galaxies (HERGs) and low-excitation radio galaxies (LERGs). All reliably classified FR-Is were found to be LERGs, while FR-IIs consist of both LERGs and HERGs (\citealt{2010A&A...509A...6B, 2012MNRAS.421.1569B}). The scheme based on optical spectra can effectively differentiate between the nuclear and host properties of FR-Is and FR-IIs (\citealt{2017A&A...598A..49C,2017A&A...601A..81C}). LERGs typically correspond to galaxies that experience inefficient accretion of hot gas, resulting in the absence of emission lines in their spectra. On the other hand, HERGs are commonly associated with galaxies that undergo rapid and efficient accretion of cold gas, which is evident from the presence of emission lines in their spectra (\citealt{2012MNRAS.421.1569B, 2016MNRAS.460....2P}). Therefore, this classification scheme proves beneficial in gaining deeper insights into the formation and evolution of radio galaxies.

In the whole FR-II-type radio galaxy sample, those with the largest linear sizes (LLS) are also known as Giant radio galaxies (GRGs; \citealt{2001A&A...371..445M, 2018ApJS..238....9K, 2020MNRAS.499...68T}). They were generally defined as those which have LLS close to or larger than 1\,Mpc (\citealt{1974Natur.250..625W, 2016A&A...594A..14P,2020A&A...641A...6P}). GRGs represent a type of the oldest radio galaxies and have an essential role in studying the AGN feedback process and duty cycle of AGN activities. However, the GRG samples are relatively small regarding to the whole radio galaxies, which is partially due to the weakness of their radio emission. Recently, with the conduction of high sensitivity and high-resolution radio surveys, such as the LOFAR Two-metre Sky Survey (LoTSS; \citealt{2017A&A...598A.104S}), the Rapid ASKAP Continuum Survey (RACS; \citealt{2020PASA...37...48M}), The MeerKAT International GHz Tiered Extragalactic Exploration Survey (MIGHTEE; \citealt{2016mks..confE...6J}), the GRG samples are increased significantly (\citealt{2020A&A...635A...5D, 2023MNRAS.525L..87C, 2021Galax...9...99A, 2021MNRAS.501.3833D}).

Moreover, the straight radio galaxies in the FRs are more readily defined and computed for radio position angles (RPAs). Analyzing the alignment of radio position angles in the orientation of radio galaxies across extensive areas of the sky can unveil the correlation between the spin axis direction of SMBH and the characteristics of cosmic filaments (\citealt{2016MNRAS.459L..36T, 2017MNRAS.472..636C}). \cite{2020A&A...642A..70O} examines the uniformity in the position angles of 7,555 double-lobed radio galaxies (a subset of FR-IIs) using data from LoTSS optical/infrared value-added catalog (\citealt{2019A&A...622A...2W}) and found that a significant deviation from the uniformity in samples' 2D analysis at angular scales of about $4^\circ$, no substantial alignment was observed in the samples' 3D analysis. \cite{2023A&A...672A.178S} investigates the distribution of radio jet position angles in radio galaxies within a 1 ${\rm deg}^2$ area of the ELAIS N1 field observed with the Giant Metrewave Radio Telescope (GMRT). The analysis reveals a significant level of spatial coherence at position angles exceeding $0.5^\circ$ and indicates a large-scale spatial coherence in terms of angular momentum.

Given the importance of the FR-II sample, we built a machine-learning-based catalog of FR-II radio galaxies from the FIRST survey. We performed a cross-match with multi-band data, and then physical properties and statistical information were presented as well. This paper is organized as follows: Sect. \ref{sect:Meth} presents the data, search strategy of the candidate of FR-IIs, and finding their optical/infrared counterparts and statistical methods. Sect. \ref{sect:results} introduces our FR-II catalog, provides an analysis and discussion of their radio and host properties, explores the GRGs included in our catalog and examines the alignment of our FR-IIs. Finally, Sect. \ref{sect:conclusion} devoted to conclusions.

Throughout the paper, we adopt a $\Lambda$CDM cosmological model with the following parameters: $H_0 = 67.4 \,{\rm km}\,{\rm s}^{-1}\,{\rm Mpc}^{-1}$, $\Omega_m = 0.315$ and $\Omega_{vac}=0.685$ (\citealt{2020A&A...641A...6P}).

\section{Data and Methods} %Methodology
\label{sect:Meth}

\subsection{The FIRST survey}
The Very Large Array (VLA) has conducted two of the most ambitious and fruitful radio sky surveys at L-band (1.4\,GHz): the Faint Images of the Radio Sky at Twenty-centimeters (FIRST; \citealt{1995ApJ...450..559B}) survey and the NRAO VLA Sky Survey (NVSS; \citealt{1998AJ....115.1693C}). The NVSS survey used the VLA D-configuration and achieved an angular resolution of $45^{\prime\prime}$ and an rms noise level of $\sim 0.45$\,mJy, while the FIRST survey adopted B-configuration and acquired a resolution of $5^{\prime\prime}$ and a typical rms noise level of 0.15\,mJy. Obviously, the FIRST survey surpasses the NVSS survey with nearly 9 times higher resolution and 3 times higher sensitivity. Therefore, the FIRST survey provides a sharper and deeper view of the radio sky, which benefits the detailed study of extragalactic radio galaxies. In this study, we employed the catalog derived from the data release FIRST-14dec17, which contains 946,432 sources (\citealt{2015ApJ...801...26H}). The FIRST-14dec17 catalog covers a total of approximately 10,575\,${\rm deg}^2$ of the radio sky (i.e. $\sim25\%$ of the entire sky), comprising 8,444\,${\rm deg}^2$ in the north Galactic cap and 2,131\,${\rm deg}^2$ in the south Galactic cap. 

The products of the FIRST survey have been extensively utilized in the past two decades for studying a diverse range of radio galaxies with different morphologies, such as Hybrid Morphology Radio Sources (HyMoRS; \citealt{2006A&A...447...63G,2022MNRAS.514.4290K}), GRGs (\citealt{2018ApJS..238....9K,2020MNRAS.499...68T}), winged radio galaxies (\citealt{2007AJ....133.2097C,2009ApJS..181..548C,2019ApJS..245...17Y,2020ApJS..251....9B}), Head-Tail (HT) radio galaxies (HTRGs, \citealt{2019A&A...626A...8M,2021ApJS..254...30P,2022ApJS..259...31S}), double-double radio galaxies (DDRGs; \citealt{2012BASI...40..121N}).

\subsection{Search strategy}
The FIRST-14dec17 catalog records the components of radio galaxies, and the association and classification of these components in radio galaxies need to be done either manually or automatically during the post-processing step (\citealt{2015MNRAS.453.2326B,2022MNRAS.514.2599S,2023RAA....23g5012L}). 
We utilize \HeTu\ (\citealt{2021SciBu..66.2145L, 2023A&C....4400728L}), a deep learning-based radio source-detector, to automatically locate and classify extended radio galaxies within the FIRST images based on their morphology. The latest version of \HeTu\ (\HeTu-v2; \citealt{2023A&C....4400728L}) has the capability to classify extended radio galaxies into four categories: FR-I, FR-II, HT, and Core-Jet (CJ). In this morphological classification scheme, we have redefined the morphology of FR-II galaxies. We adhere to the original definition of FR-II as outlined in the scheme by \cite{fanaroff1974morphology}, but only consider cases where the opening angle between the two jets is close to $180^\circ$.

The FIRST images need to be collected and pre-processed prior to source detection by \HeTu-v2. To begin with, we extracted FITS image cutouts from the FIRST Cutout Server\footnote{\url{https://third.ucllnl.org/cgi-bin/firstcutout}} centered on the positions of the sources in the FIRST-14dec17 catalog. The cutouts measure $3.96^{\prime}\times3.96^{\prime}$ ($132\times132$ pixels, where 1 pixel=1.8$^{\prime\prime}$), which ensures that extended sources fully contained within fit the cutout and maintain their morphological integrity. These cutouts were then converted into three-channel (RGB) PNG images using DS9 (\citealt{2003ASPC..295..489J}), utilizing the cool colormap and logarithmic min-max scale. 
Lastly, the \HeTu-v2 model was applied to both the FITS and PNG formats of FIRST images for source detection to build the morphological catalog (called FIRST-\HeTu), which contains fundamental physical properties of detected sources, such as their label, position, and flux density. A total of 47,582 FR-II galaxies have been detected in the FIRST-\HeTu\ catalog. Among them, 45,241 FR-IIs have been chosen for inclusion in this study. These galaxies have undergone a thorough evaluation and have been shown to have a high level of reliability with an average accuracy of 93\% (\citealt{2023A&C....4400728L}).
This paper investigates the further physical properties of the selected FR-IIs to build the definitive catalog (called FRIIRGCat).         

\subsection{Finding optical/infrared counterparts} % and properties}
We identify host galaxies of FR-II candidates compiled in this work, from the Sloan Digital Sky Survey Data Release 16 (SDSS DR16; \citealt{2020ApJS..249....3A}) and National Aeronautics and Space Administration (NASA)/Infrared Processing and Analysis Center (IPAC) Extragalactic Database (NED)\footnote{\url{https://ned.ipac.caltech.edu}}. FR-II-type radio galaxies are those with symmetric jets, it was generally observed that the host galaxies of FR-IIs are located at the symmetric center of radio structures (\citealt{2015MNRAS.453.2326B}). Accordingly, the radio-symmetric centers of FR-IIs were cross-matched with the SDSS DR16. Subsequently, we utilized the CDS X-MATCH method in TOPCAT (\citealt{2005ASPC..347...29T}) and further identified any potential optical counterparts within 15$^{\prime\prime}$, which is 3 times the angular resolution of FIRST survey. This larger radius was chosen because FR-IIs generally exhibit large structures. For any remaining galaxies without matches, we selected the source with the smallest separation available on the NED, using \astroquery\ (\citealt{2019AJ....157...98G}) with a search radius of 5$^{\prime\prime}$. During these searches, multiple counterparts may appear within the designated search range. However, following the convention of identifying host galaxies for radio sources (\citealt{2019ApJS..245...17Y}), we initially eliminate non-galactic objects and then select the one nearest to the central symmetry of FR-II as the final counterpart.
 
\begin{figure}[!ht]
   \centering
   \includegraphics[scale=0.2]{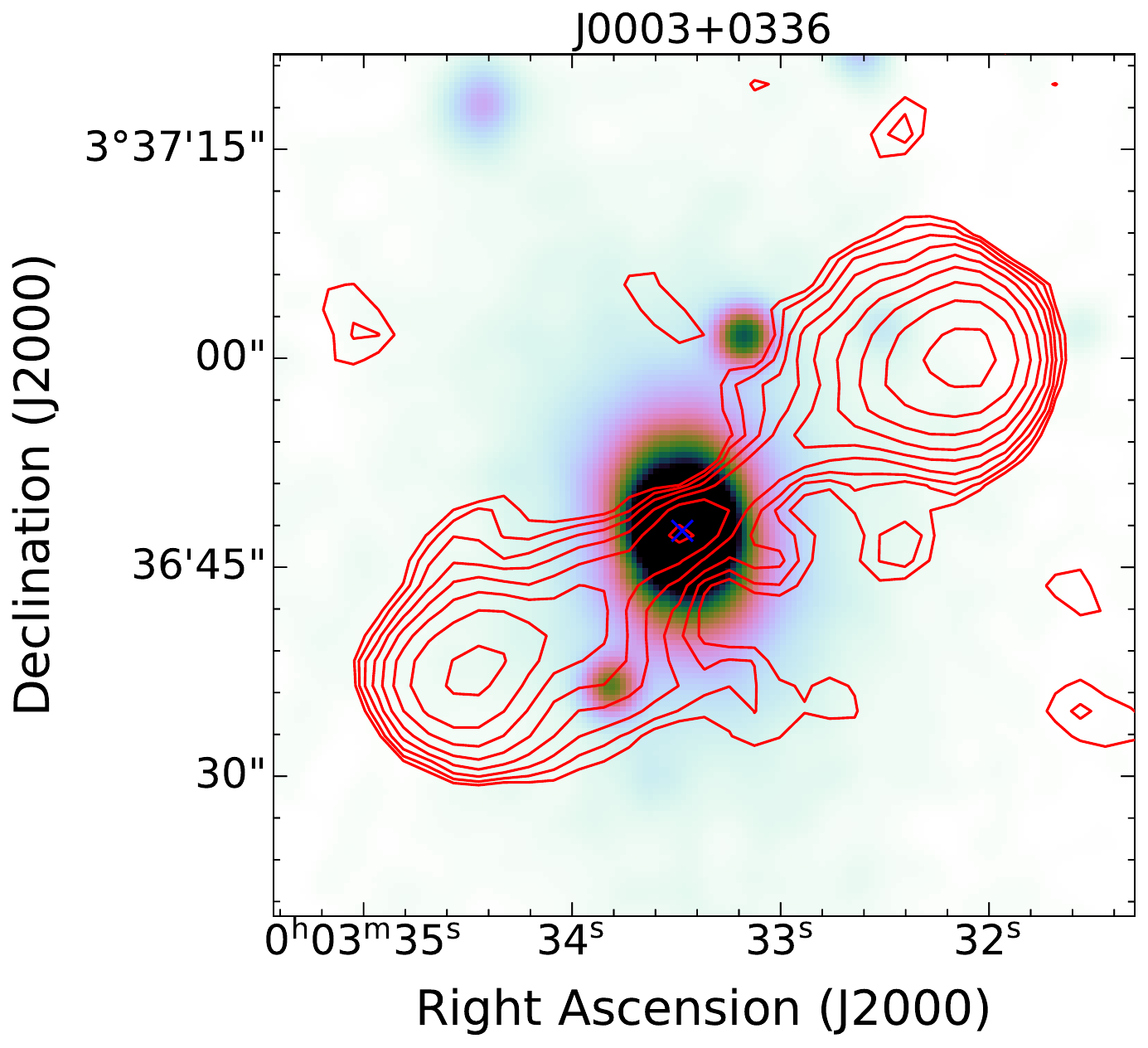}
   % \hspace{1mm}
   % \vspace{0.2mm}
   \includegraphics[scale=0.2]{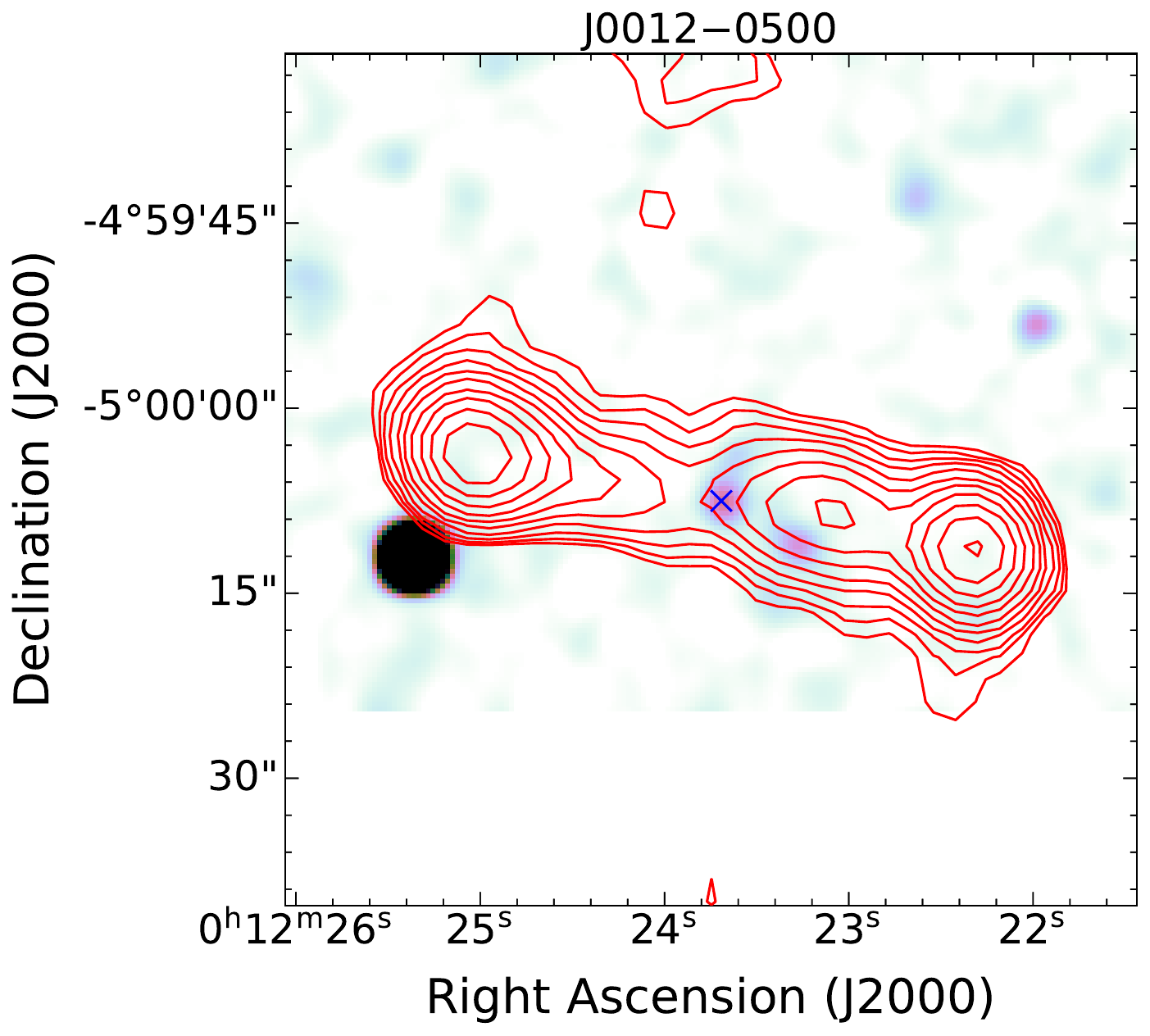}
   \includegraphics[scale=0.2]{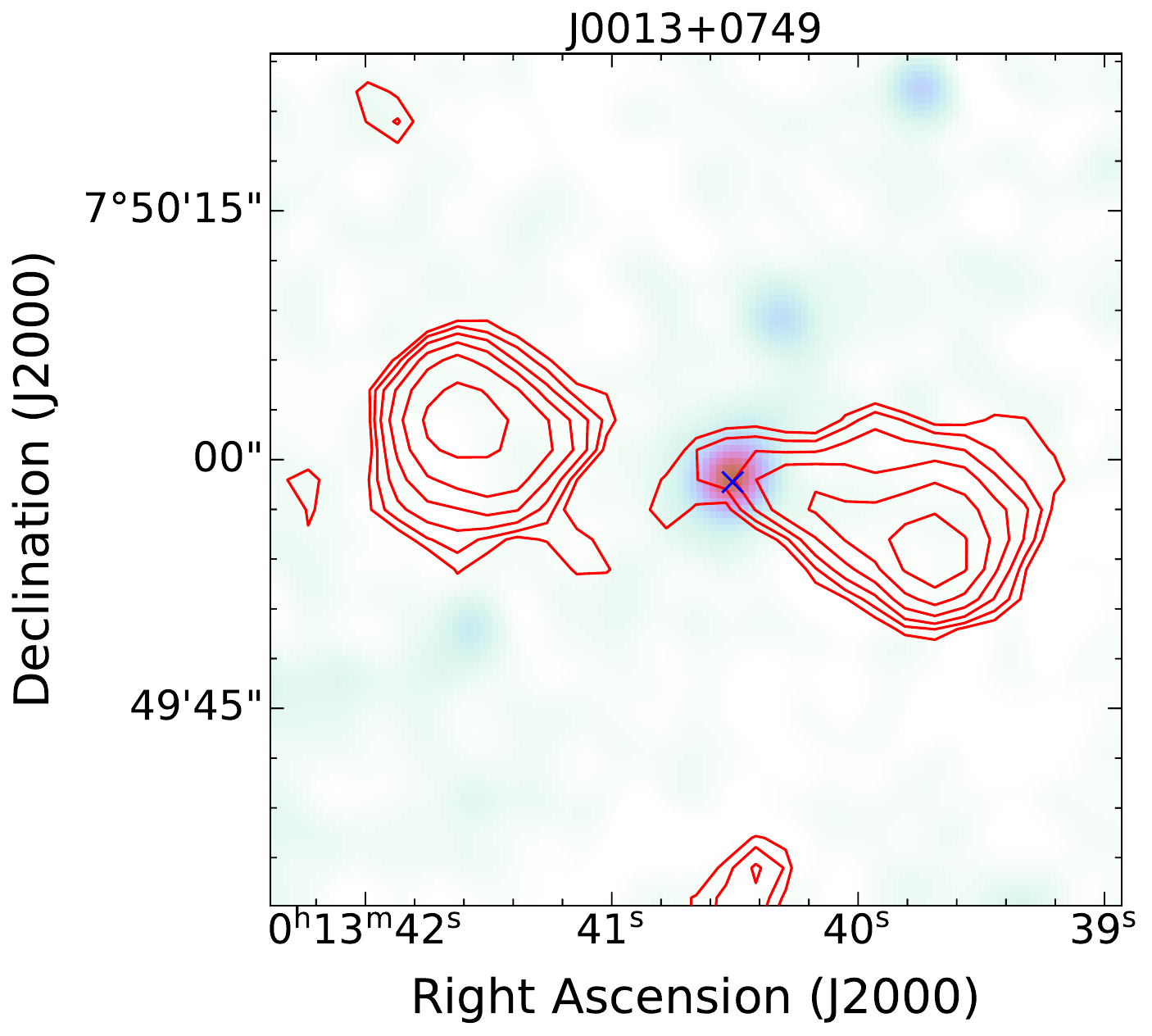}
   % \hspace{1mm}
   \includegraphics[scale=0.2]{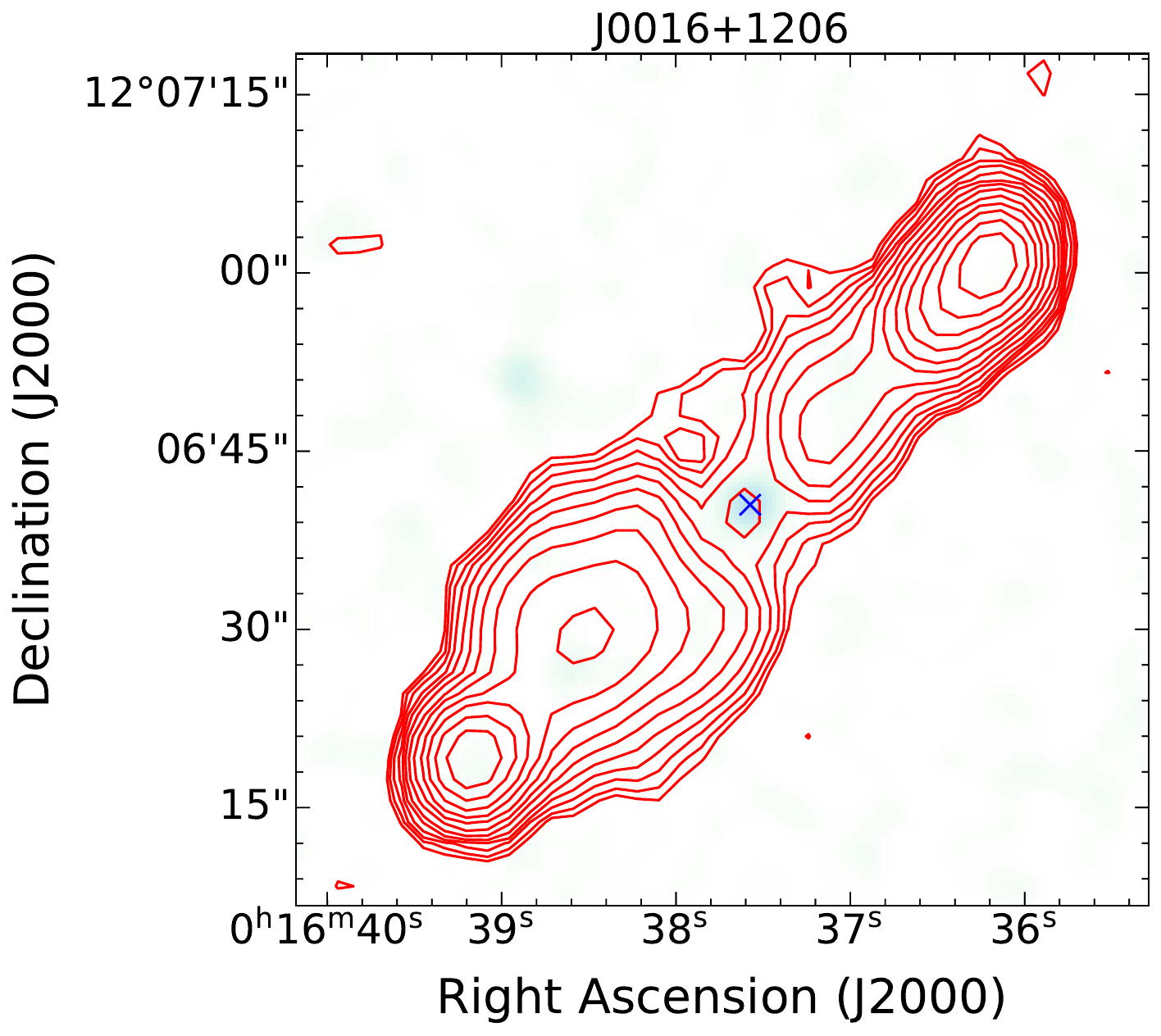}
   \includegraphics[scale=0.2]{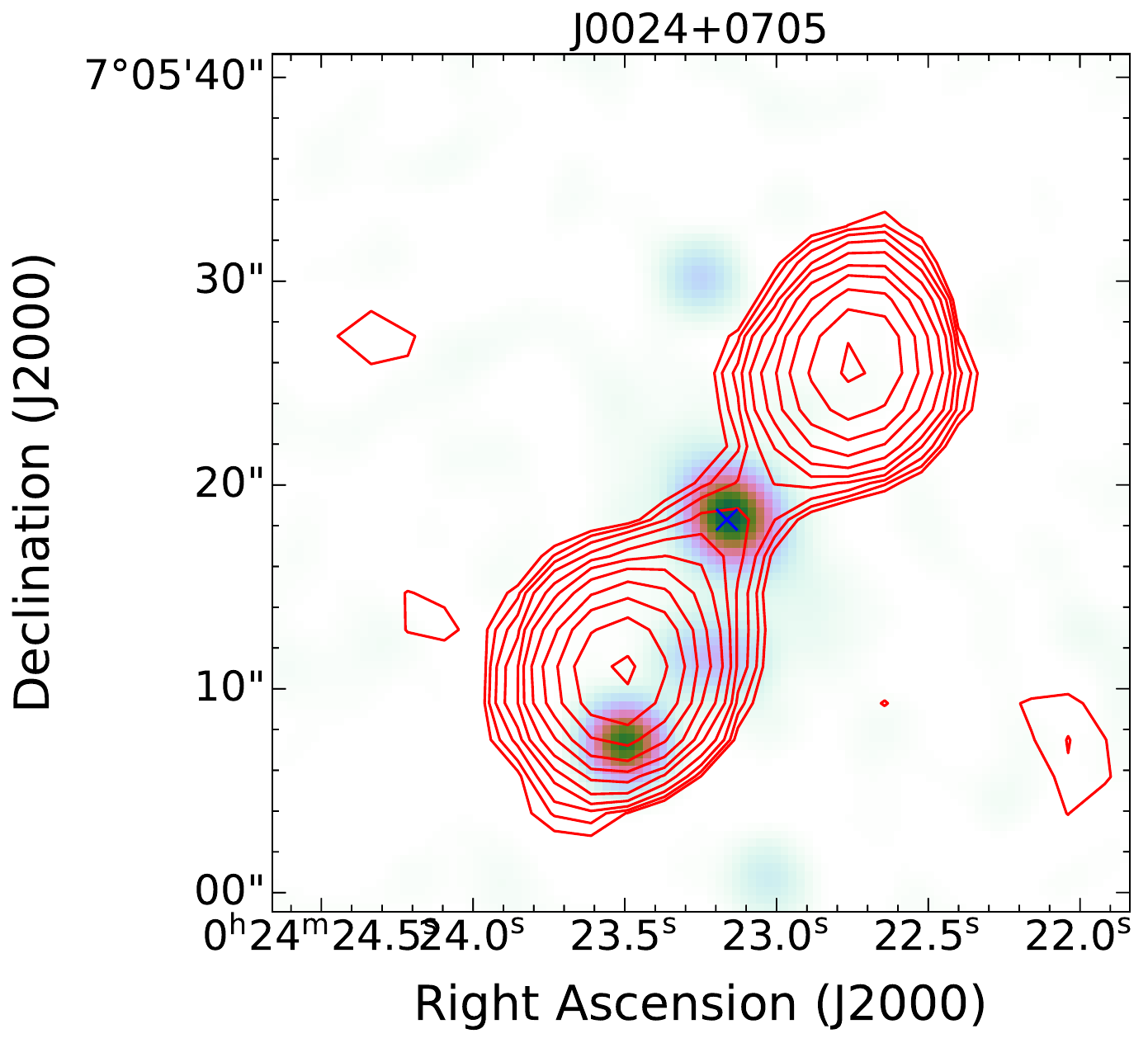}
   % \hspace{1mm}
   \includegraphics[scale=0.2]{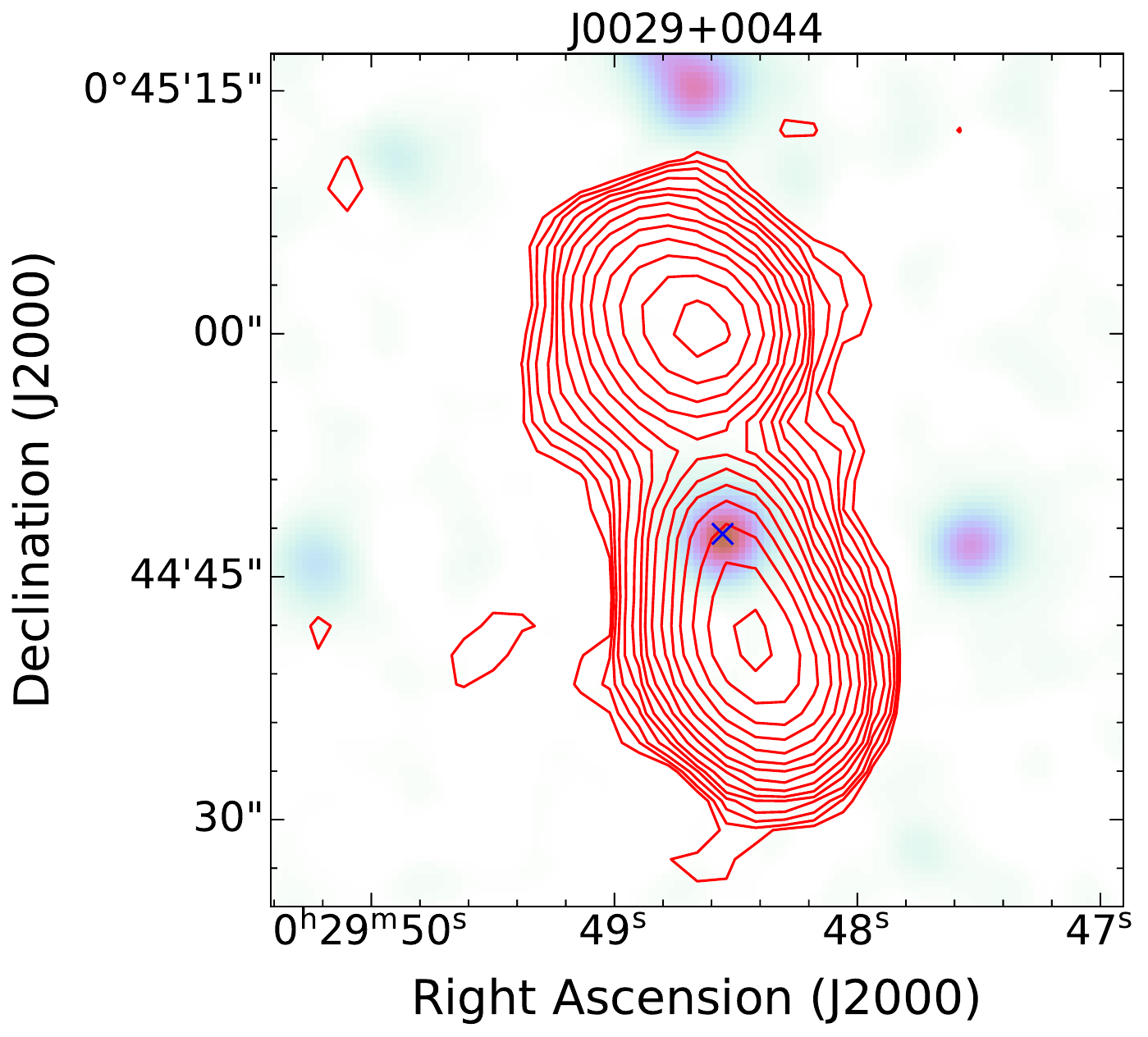}
   \includegraphics[scale=0.2]{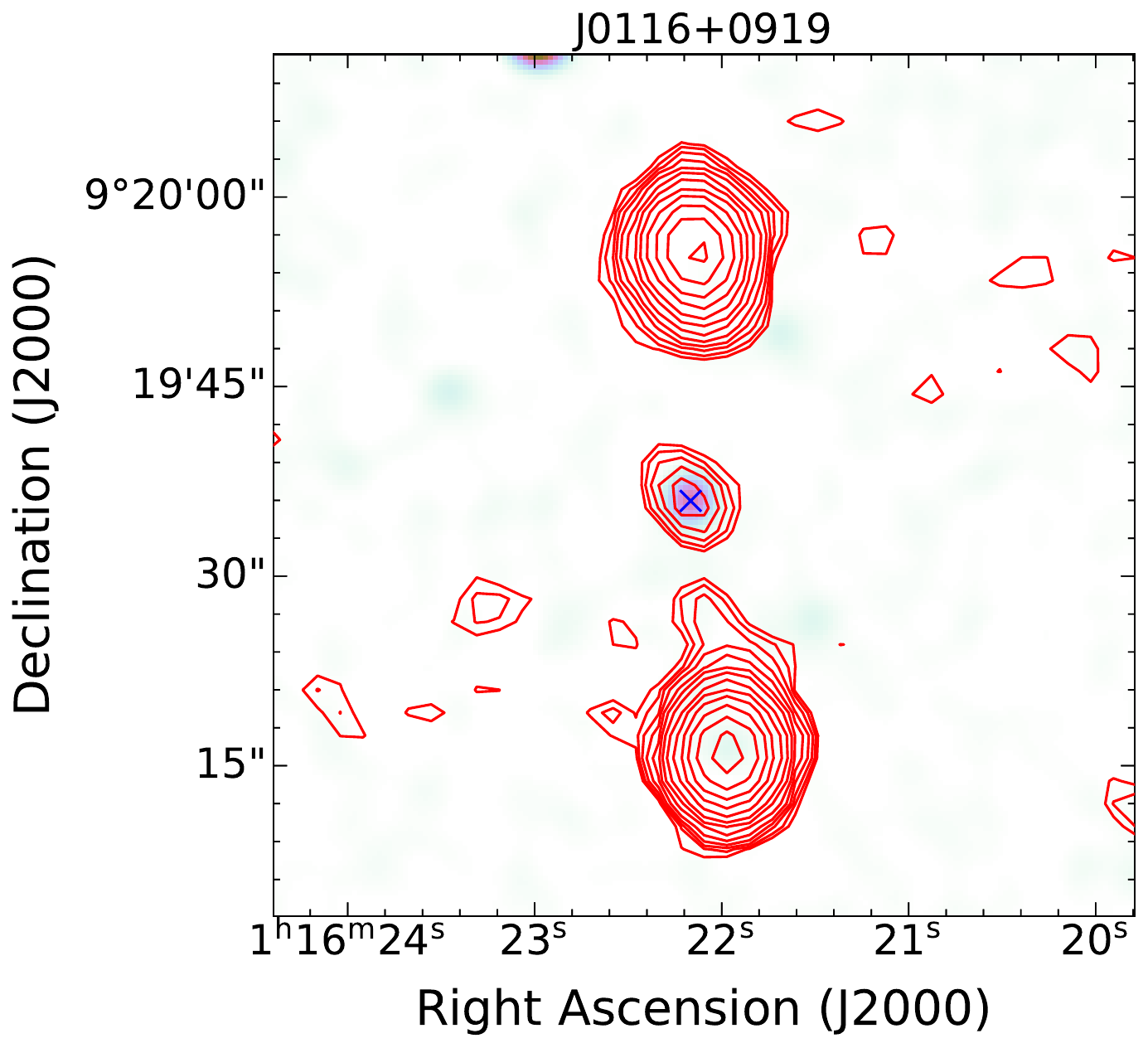}
   % \hspace{1mm}
   \includegraphics[scale=0.2]{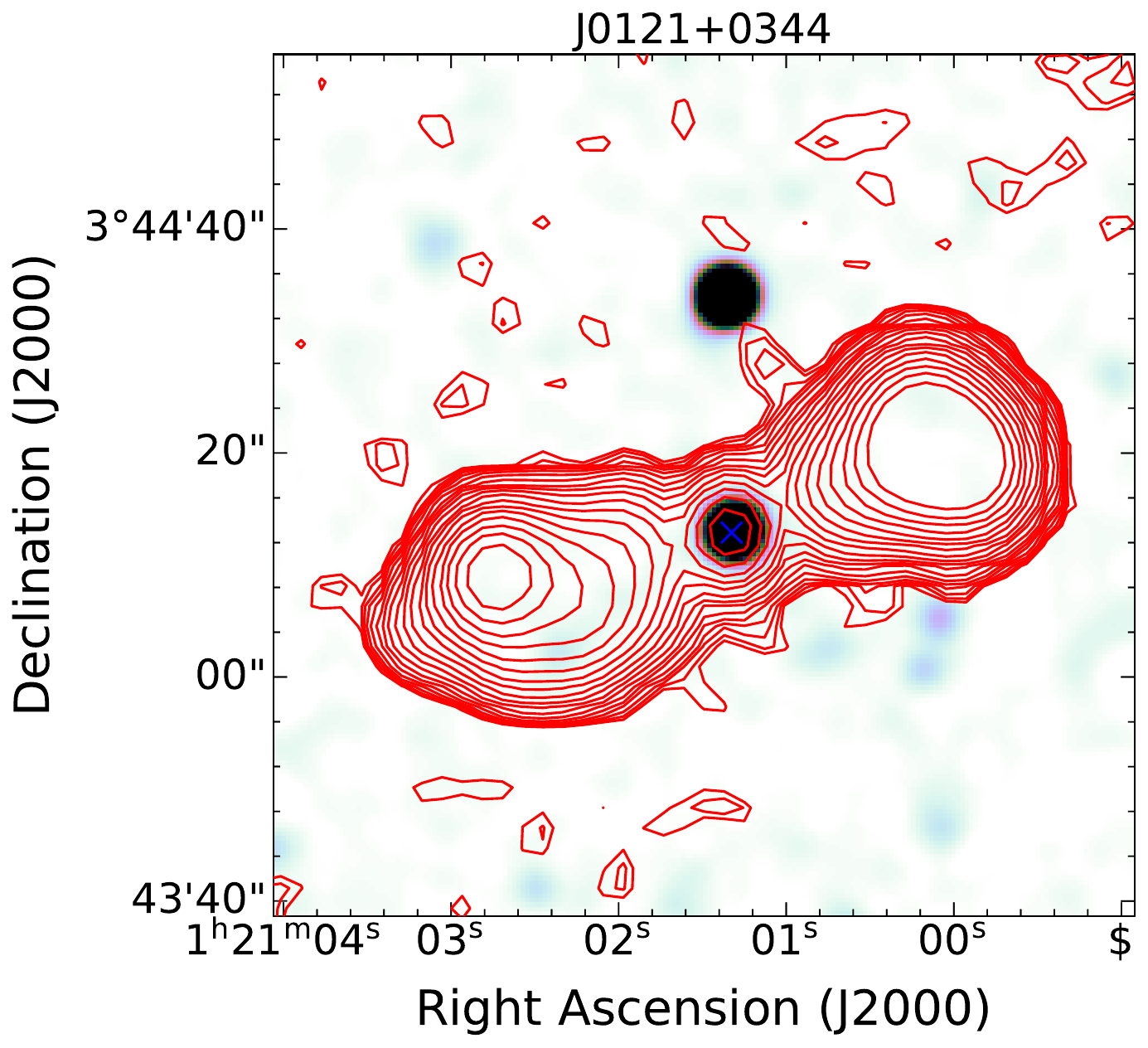}
   \includegraphics[scale=0.2]{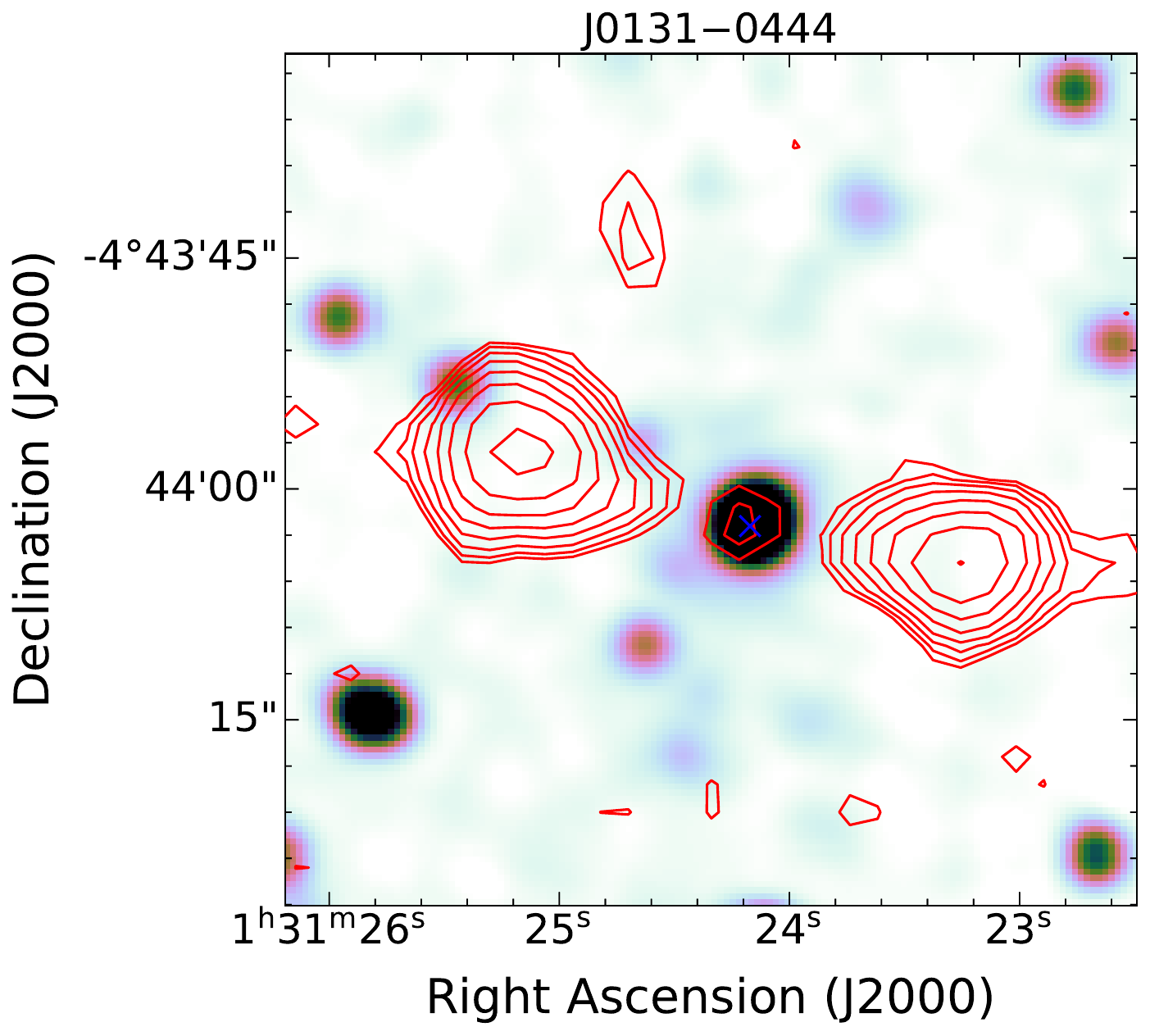}
   % \hspace{1mm}
   \includegraphics[scale=0.2]{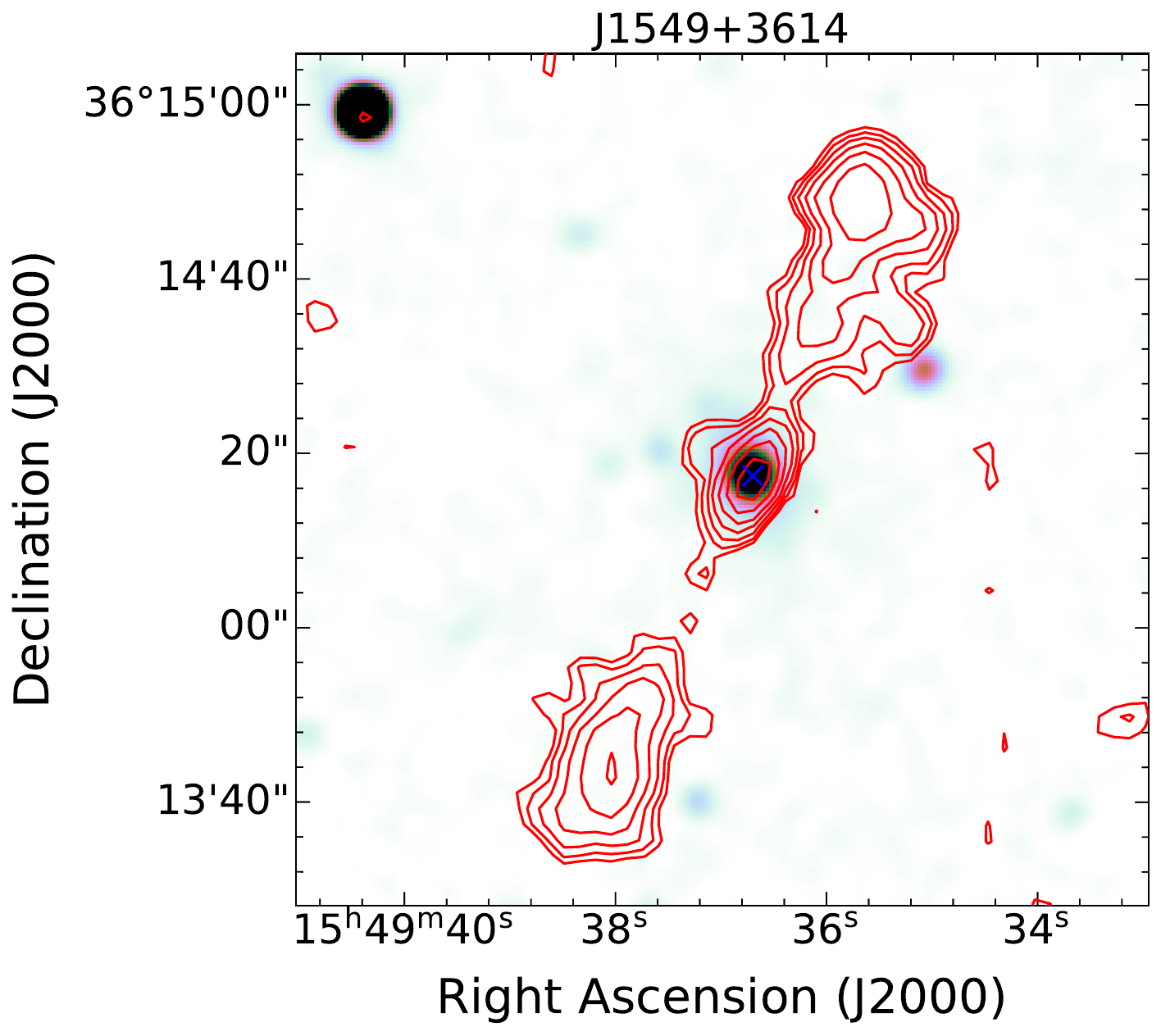}
   \includegraphics[scale=0.2]{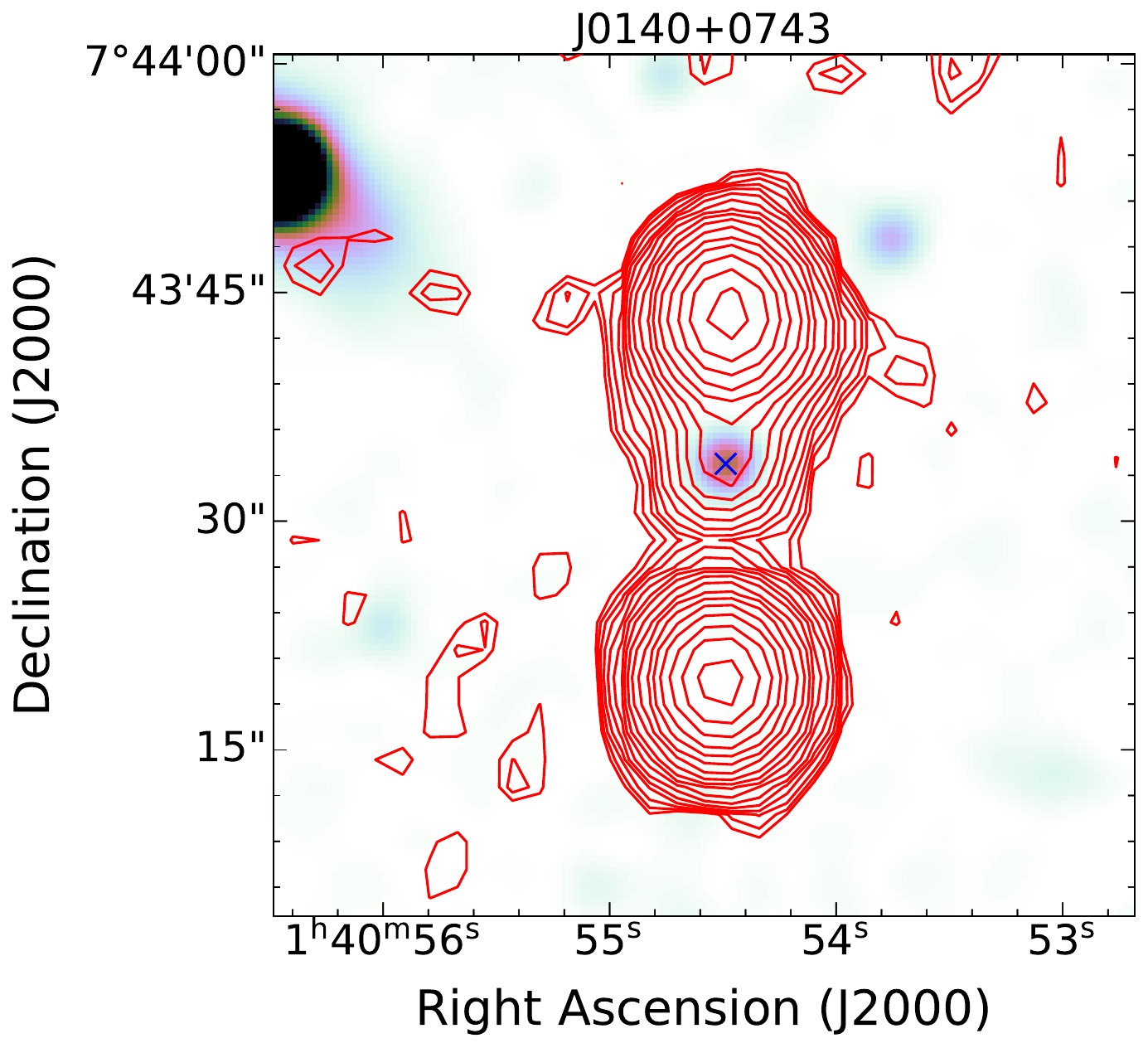}
   \includegraphics[scale=0.2]{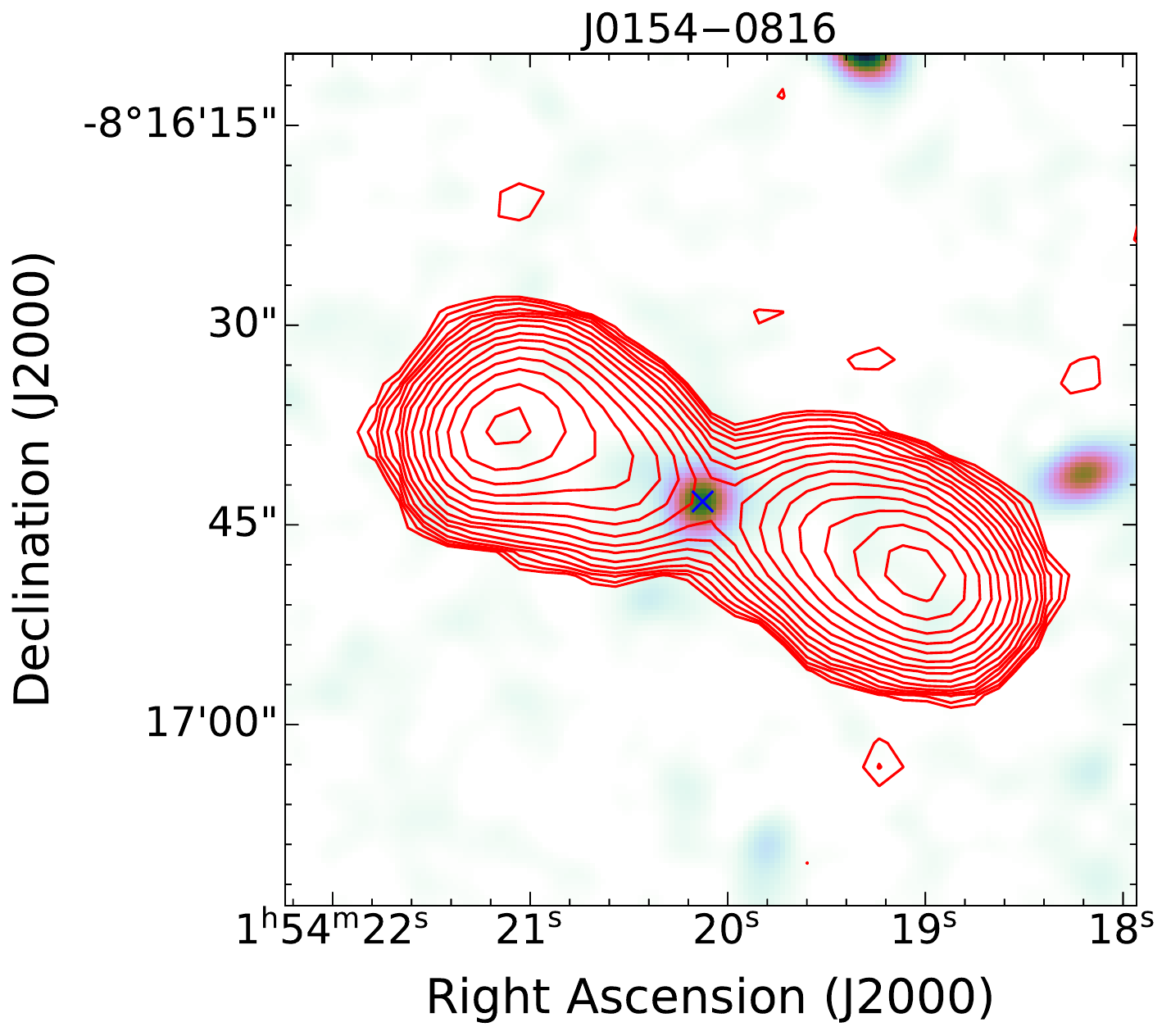}
   \caption{Radio and optical maps of a sample of 12 FR-II radio galaxies. The red color contour represents the radio emission detected in the FIRST image. The background color images are from the SDSS r-band. Contour levels are at $3\sigma\times$[1, 1.41, 2, 2.83, 4, 5.66, 8, 11.31, 16, 22.63, 32, 45.25, 64, 90.51, 128, 181.02, 256, 362.04], where $\sigma$ is the local rms noise. The blue cross is centered on the host galaxy. The background color scheme comes from CUBEHELIX (\citealt{2011BASI...39..289G}).} 
   \label{fig:radio_optical}
   \end{figure}

It is important to note that cross-identifying the host galaxies of extended radio sources has always been a challenging issue (\citealt{2018MNRAS.478.5547A}), particularly for GRGs (\citealt{2021Galax...9...99A}). Achieving a completely rigorous and accurate determination of the host galaxy may even necessitate additional optical/infrared observations. Finally, optical/infrared counterparts were identified for 41,425 (38,101 were from the SDSS DR16, 3,324 were from NED). Nearly 99\% of the results from NED are infrared counterparts, all of which were derived from the Wide-field Infrared Survey Explorer (WISE; \citealt{2011ApJ...743...34C,2011ApJS..196....4R,2013wise.rept....1C} ) data. FR-IIs that have SDSS counterparts were investigated for both their radio and optical properties. However, the remaining part of the study focused solely on examining their radio properties. In Figure~\ref{fig:radio_optical}, the FIRST image of a sample of 12 FR-II radio galaxies overlaid on the SDSS r-band image was shown, and the location of the optical counterpart of the FR-II galaxy was marked with a blue cross ($\times$) for SDSS counterparts.     

\subsection{Statistical methods for alignment of FR-IIs}
\label{sect:method-align}
The radio position angles for all FR-II galaxies were measured using the method described in \cite{2020A&A...642A..70O}. The jet position angle is estimated anticlockwisely regarding the international celestial referencing frame and the North direction. In this work, we define the radio position angle of jets as the line direction that connects two hotspots of FRIIs, i.e. neglect the bending effect of jets.  
To assess the deviation from uniformity in the alignment of radio galaxies in the sky, parallel transport and dispersion measure techniques (\citealt{2004MNRAS.347..394J, 2017MNRAS.472..636C, 2020A&A...642A..70O}) have been utilized, which take into account the effects caused by the geometry of the celestial sphere. 
Firstly, the parallel transport method has been employed to transport the ``vectors" of the radio galaxy to different locations on the celestial sphere. The specific details of this method are described below.

\begin{figure}[!ht]
   \centering
  \includegraphics[width=8cm, angle=0]{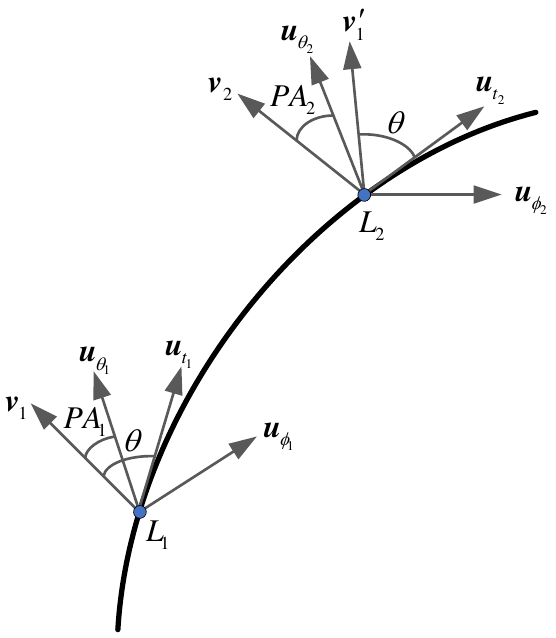}
   \caption{Schematic illustration of parallel transport. This figure illustrates the comparison between vectors ${\pmb v}_1$ and ${\pmb v}_2$. ${\pmb v}_1$ is parallel-transported to ${{\pmb v}'_1}$ along the great circle indicated by the curve from location $L_1=(\gamma_1,\theta_1, \phi_1)$ to location $L_2=(\gamma_2,\theta_2, \phi_2)$, with ${\pmb u}_{t_1}$ and ${\pmb u}_{t_2}$ being the tangent vectors to the curve at these points. The vector ${\pmb v}_1$ corresponds to the position angle $PA_1$, while the vector ${\pmb v}_2$ corresponds to the position angle $PA_2$. During parallel transport, the angle $\theta$ between the tangent vector ${\pmb u}_t$ and the vector ${\pmb v}_1$ is kept constant. This figure is re-plotted from \cite{2004MNRAS.347..394J}.} 
   \label{fig:paralle_transp}
   \end{figure}

The celestial sphere is constructed using local unit vectors (${\pmb u}_\gamma$, ${\pmb u}_\theta$, ${\pmb u}_\phi$), where each vector points towards the center of the sphere, northward and eastward on the sphere, respectively. Figure~\ref{fig:paralle_transp} shows the comparison of the radio position angles, $PA_1$ and $PA_2$, of radio galaxies 1 and 2 with locations, $L1$, and $L2$; on the celestial sphere. According to the principle in Figure~\ref{fig:paralle_transp}, the comparison of radio position angles is converted into a comparison of vectors ${\pmb v}_1$ and ${\pmb v}_2$. 

On a sphere, parallel transport enables us to establish a coordinate-invariant inner product between two vectors by translating one of them along great circle arcs that connect the two (\citealt{2017MNRAS.472..636C}). To determine the difference between ${\pmb v}_1$ and ${\pmb v}_2$, the generalized dot product between ${\pmb v}_1$ and ${\pmb v}_2$ as the dot product between the transported vector ${{\pmb v}'_1}$ and ${\pmb v}_2$ can be defined as

\begin{equation}\label{eq-dotprod}
{{\pmb v}_1} \odot {{\pmb v}_2} = {{\pmb v}'_1} \cdot {{\pmb v}_2} = \cos \left( {P{A_1} - P{A_2} + {\xi _2} - {\xi _1}} \right),
\end{equation}
where $\xi_1$ represents the angle between ${\pmb u}_{t_1}$
and ${\pmb u}_{\phi_1}$, and $\xi_2$ represents the angle between ${\pmb u}_{t_2}$
and ${\pmb u}_{\phi_2}$.

When comparing differences between radio position angles, the generalized inner product between two radio position angles can be redefined as
\begin{equation}\label{eq-inprod}
\left( {P{A_1},P{A_2}} \right) = \cos \left( {2\left( {P{A_1} - P{A_2} + {\xi _2} - {\xi _1}} \right)} \right).
\end{equation}

Considering that radio position angles range from 0 to $\pi$, the values of $(PA_1, PA_2)\in[-1,1]$, where $+1$ expresses the perfect alignment and $-1$ indicates perpendicular orientations of radio position angles (\citealt{2004MNRAS.347..394J}).

To estimate the significance of a potential alignment in FR-II radio position angles, the dispersion measure (\citealt{2004MNRAS.347..394J, 2017MNRAS.472..636C, 2020A&A...642A..70O}) will then be employed. The dispersion measure of FR-II $i$ as a function of relative RPA ($PA$) is solely determined by the difference between neighboring radio position angles, which is defined as  
\begin{equation}\label{eq-dm}
{d_{i,n}}\left( {PA} \right) = \frac{1}{n}\sum\limits_{k = 1}^n {\left( {PA,P{A_k}} \right)}, 
\end{equation}
where $n$ represents the number of nearest neighbors considered around FR-II $i$, which includes the FR-II itself. The RPA of each neighbor is denoted as $P{A_k}$. The generalized inner product ${\left( {PA,P{A_k}} \right)}$ is defined using Eq. \ref{eq-inprod}.

Then, the non-uniformity of alignment in a sample of $N$ FR-IIs is statistically determined by calculating the average of the maximum dispersion for a specified number of nearest neighbors $n$, across all $N$ FR-IIs in the sample. This measure is expressed as
\begin{equation}\label{eq-sn}
{S_n} = \frac{1}{N}\sum\limits_{i = 1}^N {{d_{i,n}}{|_{\max }}\,}, 
\end{equation}
where ${d_{i,n}}{|_{\max }}$ denotes the maximized dispersion around FR-II $i$. \cite{2004MNRAS.347..394J} confirmed that in randomly oriented samples of radio galaxies, the measure $S_n$ follows a normal distribution when the conditions $N \gg n \gg 1$ are satisfied. This property can be utilized to assess the significance of alignment across various angular scales.

Finally, by comparing the statistic $S_n$ of the data set with the statistical distribution of randomly oriented simulated samples, a significance level (SL) is determined to evaluate the rejection of the null hypothesis that the galaxy samples exhibit random orientation. Therefore, the measurements of $S_n$ can be transformed into one-tailed significance level when analyzing multiple values of $n$:
\begin{equation}\label{eq-SL}
{\rm SL} = 1 - \Phi \left( {\frac{{{S_n} - {{\left\langle {{S_n}} \right\rangle }_{\rm MC}}}}{{{\sigma _n}}}} \right),
\end{equation}
where $\Phi$ represents the cumulative normal distribution function, and ${{\left\langle {{S_n}} \right\rangle }_{\rm MC}}$ signifies the expected value for $S_n$ in the absence of alignment. This expected value is determined through Monte Carlo simulations. 
%\cite{2017MNRAS.472..636C} and \cite{2020A&A...642A..70O} have verified that 
A lower minimum significance level indicates stronger evidence for rejecting the null hypothesis, suggesting a higher degree of alignment for radio position angles.

Furthermore, if the redshift of the FR-IIs is known, the above approach can be extended to identify nearest neighbors in 3D space. This enables the detection of potential alignment effects and the evaluation of the dependence of $S_n$ on a physical scale.

\section{Results and Discussion}
\label{sect:results}
\subsection{The new FR-II radio galaxies catalog}
This new FR-II radio galaxies catalog (FRIIRGcat\footnote{\url{https://drive.google.com/file/d/19m_ma-2fFIWVZ8WJphXxr5W_HXkyIAeX/view?usp=drive_link}}) contains a total of 45,241 candidates. With the available radio and optical data, the LLS, RPA, spectral indices, radio luminosities, ${\rm [O\,III]}$ luminosities, and other host properties of all FR-II candidates have been calculated. Based on these results, the statistical properties and the uniformity of the alignment of these candidates have also been studied. The first 50 rows of FR-II candidates in FRIIRGcat are listed in Table \ref{tab:FRIIcand}. The optical/infrared counterparts of FR-II sources are listed in the 4th column of Table \ref{tab:FRIIcand}, along with the name of the catalog/telescope from which they were identified. The positions of the optical/infrared counterparts were used as the positions of FR-IIs and are presented in the 2nd and 3rd columns of Table \ref{tab:FRIIcand}. When no clear optical counterparts are available, the central positions from the FIRST-\HeTu\ catalog are utilized as the positions for the FR-II sources, which are indicated as `HeTu'.

The redshift ($z$) values of the candidates were estimated from the SDSS and NED catalogs and listed in column 5. The redshifts were measured through photometric or spectroscopic measurements. For each SDSS counterpart, we first utilized the spectroscopic redshift queried from the spectral data of SDSS DR16 as the redshift. In cases where the spectroscopic redshift was not available, we then adopted the photometric redshift queried from the `Photoz' table in the SDSS DR16 database as the redshift value. The redshifts obtained from NED are determined by the type code in the `Redshift Flag' column in the search results, indicating whether they are spectroscopic or photometric redshifts. Redshifts were successfully determined for 20,147 out of 45,241 (44.5\%) FR-IIs, where the number of photometric and spectroscopic redshifts is 7,261 and 12,886, respectively. Most photometric redshifts (12,400) were taken from SDSS DR16. These redshifts were estimated using a robust fitting method that considered the nearest neighbors in a reference set (\citealt{2016MNRAS.460.1371B}). The estimated error for these redshifts is at the level of (substantially smaller than) 0.1 (\citealt{2005astro.ph..8564S,2007ApJS..172..634A,2016MNRAS.460.1371B}), resulting in large uncertainties in photometric redshift during 3D alignment analysis (\citealt{2017MNRAS.472..636C}). Among all candidates, J1053+0604 has the highest redshift value, 5.01320$\pm$0.00167, and J1157+3220 has the lowest redshift value, of 0.01088$\pm$0.00001. Four candidates have redshift values greater than 4: J1053+0604, J1239+1005, J1159+0347 and J1506+4504. The redshifts of these four candidates were taken from SDSS DR13 (\citealt{2017ApJS..233...25A}). 

\begin{figure}[!ht]
   \centering
  \includegraphics[width=12cm, angle=0]{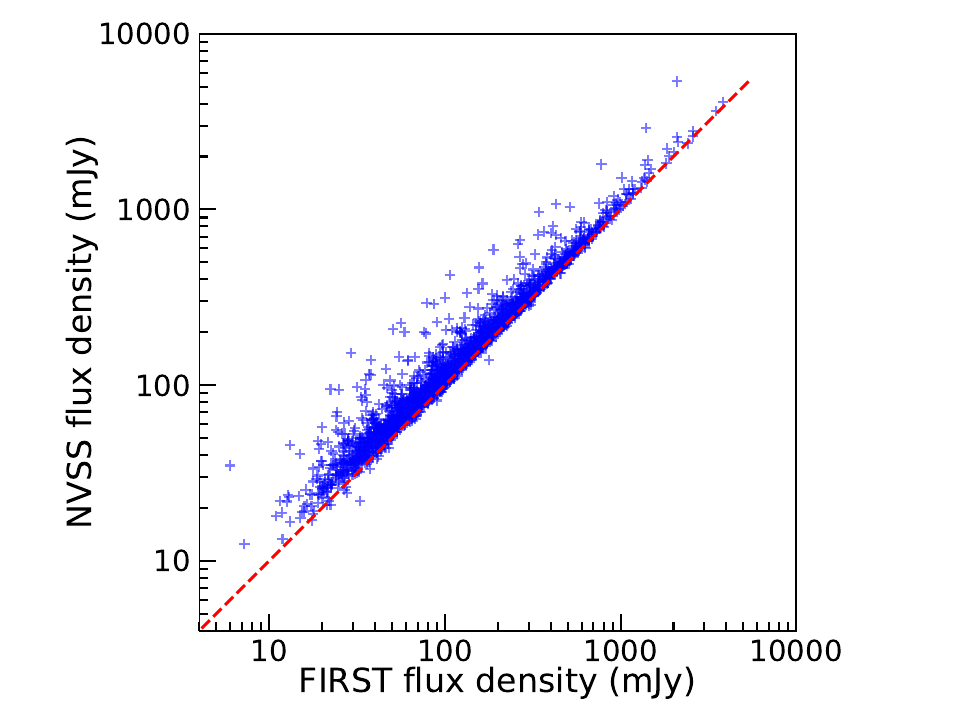}
   \caption{Flux densities measured from the FIRST and NVSS were shown for 5\% (2,380) of complete FR-II candidates presented in this paper. The red dotted line represented the 1:1 ratio
between flux densities from the FIRST and NVSS surveys.} 
   \label{fig:FIRSTvsNVSS}
   \end{figure}

Columns 6, 7 and 8 represent the values for the RPA, largest angular size (LAS) and LLS of all FR-IIs, respectively. The LAS was determined by measuring the distance between the endpoints of the two active radio lobes, as measured in the FIRST map. In column 9, we chose to measure the radio-integrated flux density at 1.4 GHz of FR-IIs using both NVSS data and FIRST data. In Figure~\ref{fig:FIRSTvsNVSS}, the distribution of flux density for a random selection of FR-II radio galaxies is shown, comparing the FIRST and NVSS measurements. This comparison confirms that a majority (94.6\%) of the flux densities derived from the NVSS images are notably higher than those obtained from the FIRST images. This is because the NVSS has more accurate fluxes for extended radio galaxies, allowing it to detect low-surface-brightness objects that may be missed by the FIRST survey. Additionally, this can be attributed to the higher resolution of the FIRST and its lack of antennas for short spacing. For these FR-IIs, flux density measurements used NVSS flux densities. However, a small minority (5.4\%) of NVSS flux densities are smaller than FIRST fluxes. In such cases, the flux density measurements used the FIRST fluxes. The brightest candidate for FR-II is J1229+0203, with a flux ($S_{1.4}$) of 55\,152 mJy. Conducting higher-resolution observations will facilitate the examination of various aspects of this candidate, including its dynamics.

%top 50 rows
%\usepackage{longtable}	
%\begin{longtable}{ccccccccccc}
%\end{longtable}
\begin{landscape}
% \begin{table}[!ht]
% \begin{sidewaystable}%[!htb]
% \bc
\begin{minipage}[]{80mm}
%\captionsetup{justification=centering} 
%\caption{The top 50 FR-II candidates in FRIIRGcat}
\end{minipage}
\centering
\setlength{\tabcolsep}{2pt}
\small
% \begin{center}
 \begin{longtable}{ccccccccccccccccccc} %tabular 
  \caption{The first 50 rows of FR-II candidates in FRIIRGcat \label{tab:FRIIcand}} \\ 
  \hline \noalign{\smallskip}
Name&R.A. & Decl. & Ref.&$z$ &RPA & LAS & LLS & $S_{1.4}$& $S_{3}$ & $\alpha_{1.4}^{3}$ & ${ \rm log}(L_{\rm rad})$ & Dn4000 & $C_{\rm r}$& $M_{\rm r}$ & log($M_{\rm BH})$ & log($L_{\rm [O\,III]}$)&Class    \\
& (J2000.0)&(J2000.0)&&&(deg)&(arcsec)&(kpc)&(mJy)& (mJy)&& ${\rm{(W}}\,{{\rm Hz}^{ - 1}})$ &&&(mag) &($M_{\odot}$)&(${\rm erg}\,{\rm s}^{-1}$)  & \\
  \hline\noalign{\smallskip}
% J0757+1904& &2012-02-26&&2700&20.40&19.45&19.00&18.53&17.88&17.02&xx && && &&\\
J0000$+$0441& 00 00 01.35& $+$04 41 29.6& SDSS& 0.58$^{s}$&20.0& 21& 143& 22 & 12 & 0.89& 26.49 &$-$&2.19&$-$22.35&8.41&$-$&$-$ \\
J0000$+$0957& 00 00 02.90& $+$09 57 00.7& SDSS& $-$       &168.1& 35& $-$& 306& 146& 0.97& $-$ &$-$&$-$&$-$&$-$&$-$&$-$ \\
J0000$+$0958& 00 00 06.63& $+$09 58 51.0& SDSS& $-$       &26.6& 24& $-$& 306& $-$& $-$ & $-$ &$-$&$-$&$-$&$-$&$-$&$-$ \\
J0000$-$0317& 00 00 06.65& $-$03 17 54.8& SDSS& $-$       &151.7& 27& $-$& 40 & 18 & 1.06& $-$ &$-$&$-$&$-$&$-$&$-$&$-$ \\
J0000$-$0343& 00 00 12.23& $-$03 43 01.0& SDSS& 1.37$^{s}$&65.2& 26& 225& 60 & 40 & 0.52& 27.68 	&$-$&2.31&$-$25.6&$-$&$-$&$-$ \\
J0000$+$1002& 00 00 15.86& $+$10 02 24.7& SDSS& 0.75$^{s}$&141.9& 32& 242& 71 & 37 & 0.86& 27.25 &$-$&2.71&$-$21.57&7.16&$-$&$-$ \\
J0000$-$0236& 00 00 16.38& $-$02 36 48.4& SDSS& 0.51$^{p}$&125.2& 37& 236& 63 & 33 & 0.87& 26.81 &$-$&$-$&$-$&$-$&$-$&$-$ \\
J0000$-$0851& 00 00 17.37& $-$08 51 23.7& SDSS& 1.25$^{s}$&108.5& 23& 197& 23 & 10 & 1.17& 27.40 &$-$&2.06&$-$26.48&$-$&$-$&$-$ \\
J0000$-$0808& 00 00 17.98& $-$08 08 55.5& SDSS& 0.68$^{p}$&39.1& 37& 270& 41 & 16 & 1.27& 27.01 &$-$&$-$&$-$&$-$&$-$&$-$ \\
J0000$+$1229& 00 00 24.06& $+$12 29 53.2& SDSS& 0.66$^{p}$&23.1& 69& 497& 109& 46 & 1.14& 27.37 &$-$&$-$&$-$&$-$&$-$&$-$ \\
J0000$+$0023& 00 00 25.56& $+$00 23 28.9& SDSS& $-$       &154.5& 42& $-$& 48 & 20 & 1.18& $-$   &$-$&$-$&$-$&$-$&$-$&$-$ \\
J0000$-$1121& 00 00 30.11& $-$11 21 19.2&WISEA& 0.10$^{s}$&147.2& 30& 57 & 81 & 47 & 0.72& 25.33 &$-$&$-$&$-$&$-$&$-$&$-$ \\
J0000$+$0411& 00 00 40.35& $+$04 11 31.7& SDSS& $-$       &144.4& 31& $-$& 73 & 38 & 0.86& $-$   &$-$&$-$&$-$&$-$&$-$&$-$ \\
J0000$-$0721& 00 00 40.12& $-$07 21 32.1& SDSS& $-$       &73.3&132& $-$& 135& 48 & 1.35& $-$   &$-$&$-$&$-$&$-$&$-$&$-$ \\
J0000$+$1435& 00 00 42.35& $+$14 35 45.2& SDSS& 0.62$^{s}$&161.5& 23& 161& 15 & 8  & 0.84& 26.38 &$-$&2.23&$-$22.25&$-$&$-$&$-$ \\
J0000$-$1013& 00 00 42.85& $-$10 13 37.1& SDSS& 0.66$^{p}$&117.6& 47& 338& 191& 95 & 0.92& 27.56 &$-$&$-$&$-$&$-$&$-$&$-$ \\
J0000$-$0041& 00 00 43.69& $-$00 41 30.1& SDSS& $-$       &108.4& 23& $-$& 17 & 6  & 1.47& $-$ &$-$&$-$&$-$&$-$&$-$&$-$ \\
J0000$+$1216& 00 00 44.11& $+$12 16 10.6& SDSS& $-$       &26.5& 28& $-$& 58 & 14 & 1.88& $-$ &$-$&$-$&$-$&$-$&$-$&$-$ \\
J0000$-$0050& 00 00 48.76& $-$00 50 32.7& SDSS& $-$       &116.6& 40& $-$& 22 & 17 & 0.31& $-$ &$-$&$-$&$-$&$-$&$-$&$-$ \\
J0000$+$0725& 00 00 52.78& $+$07 25 23.6& SDSS& 0.59$^{p}$&66.8& 14& 96 & 19 & 13 & 0.5 & 26.36 &$-$&$-$&$-$&$-$&$-$&$-$ \\
J0000$-$0110& 00 00 52.96& $-$01 10 20.5& SDSS& 0.19$^{s}$&16.7& 19& 62 & 11 & 7  & 0.54& 25.05 &1.92&2.9&$-$22.21&8.30&39.69&LERG \\
J0000$+$0210& 00 00 53.75& $+$02 10 18.6& SDSS& 0.46$^{p}$&6.0& 34& 205& 67 & 18 & 1.78& 26.88 &$-$&$-$&$-$&$-$&$-$&$-$ \\
J0000$+$0325& 00 00 54.84& $+$03 25 36.3& SDSS& 0.54$^{s}$&98.7& 24& 157& 36 & 18 & 0.91& 26.63 &$-$&3.21&$-$22.58&9.16&$-$&$-$ \\
J0001$+$0403& 00 01 02.64& $+$04 03 12.0&WISEA& $-$       &100.3& 40& $-$& 72 & 39 & 0.80& $-$   &$-$&$-$&$-$&$-$&$-$&$-$\\
J0001$-$0939& 00 01 09.17& $-$09 39 57.1&WISEA& $-$       &56.8& 62& $-$& 17 & 8  & 1.14& $-$   &$-$&$-$&$-$&$-$&$-$&$-$ \\
J0001$+$0438& 00 01 12.10& $+$04 38 55.0& SDSS& $-$       &150.2& 29& $-$& 28 & 17 & 0.71& $-$   &$-$&$-$&$-$&$-$&$-$&$-$ \\
J0001$+$0820& 00 01 14.98& $+$08 20 29.1& SDSS& 0.54$^{s}$&30.1& 129& 846&404& 155& 1.26& 27.75 &$-$&2.25&$-$23.57&$-$&$-$&$-$ \\
J0001$+$0545& 00 01 16.09& $+$05 45 39.3& SDSS& 0.52$^{s}$& 71.6& 17& 109& 17& 13& 0.34& 26.16& $-$& 1.62& $-$20.95& 7.26& $-$& $-$ \\
J0001$+$1122& 00 01 16.58& $+$11 22 21.2& SDSS& 0.38$^{p}$& 118.1& 61& 328& 40& 23& 0.77& 26.30 &$-$&$-$&$-$&$-$&$-$&$-$\\
J0001$-$0153& 00 01 18.81& $-$01 53 27.5& SDSS& $-$& 120.3& 25& $-$& 18& 8& 1.15& $-$ &$-$&$-$&$-$&$-$&$-$&$-$ \\
J0001$+$0324& 00 01 19.74& $+$03 24 40.6& SDSS& $-$ & 149.7& 25& $-$& 18& 12& 0.54& $-$ &$-$&$-$&$-$&$-$&$-$&$-$\\
J0001$+$0853& 00 01 24.81& $+$08 53 25.7& SDSS& 1.05$^{s}$& 72.6& 30& 250& 38& 25& 0.57& 27.24& $-$& 2.17& $-$24.0&$-$&$-$&$-$ \\
J0001$+$0936& 00 01 26.00& $+$09 36 32.0& WISEA& $-$& 23.0& 78& $-$& 143& 71& 0.93& $-$&$-$&$-$&$-$&$-$&$-$&$-$ \\
J0001$+$0114& 00 01 26.10& $+$01 14 32.9& SDSS& $-$ & 156.8& 41& $-$& 70& 33& 1.01& $-$ &$-$&$-$&$-$&$-$&$-$&$-$\\
J0001$-$0208& 00 01 27.28& $-$02 08 41.6& SDSS& $-$ & 16.7& 19& $-$& 37& 26& 0.50& $-$ & 2.36& $-$ &$-$&$-$&$-$&$-$\\
J0001$+$0148& 00 01 30.19& $+$01 48 16.1& SDSS& $-$ & 116.6& 16& $-$& 26& 12& 1.09 &$-$ &$-$&$-$&$-$&$-$&$-$&$-$\\
J0001$-$1057& 00 01 33.18& $-$10 57 45.4& SDSS& $-$ & 18.4& 125& $-$& 114& 41& 1.35& $-$ &$-$&$-$&$-$&$-$&$-$&$-$\\
J0001$-$1114& 00 01 34.15& $-$11 14 12.8& WISEA& $-$ & 21.0& 25& $-$& 9& 5& 0.95& $-$ &$-$&$-$&$-$&$-$&$-$&$-$\\
J0001$-$0851& 00 01 34.90& $-$08 51 54.7& SDSS& 0.18$^{s}$& 114.8& 26& 82& 62& 34& 0.78& 25.77& 1.81& 3.62& $-$22.6& 8.31& 40.3& LERG\\
J0001$+$0034& 00 01 36.27& $+$00 34 03.7& SDSS& 0.89$^{p}$& 123.7& 26& 208& 27& 16& 0.69& 26.96&$-$&$-$&$-$&$-$&$-$&$-$\\
J0001$-$0410& 00 01 40.86& $-$04 10 30.2& SDSS& 0.96$^{p}$& 164.7& 21& 171& 10& 5& 1.01& 26.70 &$-$&$-$&$-$&$-$&$-$&$-$\\
J0001$+$0711& 00 01 43.32& $+$07 11 57.1& SDSS& $-$ & 47.3& 32& $-$& 17& 12& 0.45 & $-$ &$-$&$-$&$-$&$-$&$-$&$-$\\
J0001$-$0141& 00 01 45.09& $-$01 41 33.4& SDSS& $-$ & 157.4& 23& $-$& 22& 13& 0.73& $-$ &$-$&$-$&$-$&$-$&$-$&$-$\\
J0001$-$1034& 00 01 49.93& $-$10 34 49.0& SDSS& $-$ & 72.9& 24& $-$& 29& 16& 0.77& $-$ &$-$&$-$&$-$&$-$&$-$&$-$\\
J0001$+$0204& 00 01 54.76& $+$02 04 49.7& SDSS& 0.41$^{s}$& 16.4& 64& 361& 309& 135& 1.09& 27.31& $-$& 2.88& $-$22.28& 8.56 & $-$& $-$ \\
J0001$-$0704& 00 01 56.01&$-$07 04 02.0& SDSS& 0.44$^{p}$& 157.7& 43& 253& 60& 27& 1.07& 26.67 &$-$&$-$&$-$&$-$&$-$&$-$\\
J0001$-$0944& 00 01 59.01& $-$09 44 58.9& SDSS& $-$ & 149.0& 31& $-$& 19& 10& 0.84& $-$ &$-$&$-$&$-$&$-$&$-$&$-$\\
J0002$+$1433& 00 02 02.51& $+$14 33 33.9& SDSS& 0.49$^{p}$& 50.6& 40& 250& 92& 53& 0.73& 26.91 &$-$&$-$&$-$&$-$&$-$&$-$\\
J0002$-$0002& 00 02 02.33& $-$00 02 44.7& SDSS& $-$ & 28.8& 41& $-$& 22& 9& 1.16& $-$&$-$&$-$&$-$&$-$&$-$&$-$\\
J0002$+$1106& 00 02 04.99& $+$11 06 54.4& SDSS& 0.40$^{p}$& 90.0& 22& 122& 53& 27& 0.87& 26.49 &$-$&$-$&$-$&$-$&$-$&$-$ \\
  \noalign{\smallskip} \hline
\end{longtable}
% \end{center}
% \end{tabular}
\tablecomments{1.4\textwidth}{Column 1: short name. Column 2: R.A. (J2000, hh:mm:ss). Column 3: Decl. (J2000, dd:mm:ss). Column 4: reference for the position. Column 5: redshift, $^{p}$represents photometric redshifts,
$^{s}$represents spectroscopic redshifts. Column 6: radio position angle. Column 7: largest angular size. Column 8: projected largest linear size. Column 9: integrated flux density at 1.4 GHz. Column 10: integrated flux density at 3 GHz. Column 11: spectral index between 1.4 GHz and 3 GHz. Column 12: logarithm luminosity at 1.4 GHz. Column 13: 4000-\AA break index. Column 14: concentration index. Column 15: r-band absolute magnitude. Column 16: logarithm black hole mass. Column 17: logarithm [O III] line luminosity. Column 18: LERG/HERG classification result.}
% \end{table}
\end{landscape}
% \end{sidewaystable}

The two-point spectral indexes ($\alpha_{1.4}^3$) for all FR-IIs between 1.4 GHz (NVSS or FIRST) and 3 GHz from the Karl G. Jansky VLA Sky Survey (VLASS; \citealt{2020PASP..132c5001L}) were listed in column 11. The VLASS flux densities listed in column 10 were measured from the cutout VLASS images, which were obtained by utilizing the cutout core command-line tool\footnote{\url{https://github.com/CIRADA-Tools/cutout_provider_core}} within the Canadian Initiative for Radio Astronomy Data Analysis (CIRADA). In FRIIRGcat, spectral index ($\alpha_{1.4}^3$) was available for 44,763 FR-IIs. The remaining 478 FR-IIs due to the VLASS fluxes were not detectable in the CIRADA VLASS catalog at that time. The radio luminosity at 1.4 GHz for all FR-II candidates was computed in column 12.

The optical properties of FR-IIs were obtained from the SDSS DR16 spectroscopic catalog and the Max Planck Institute for Astrophysics and Johns Hopkins University (MPA-JHU, \citealt{2004MNRAS.351.1151B}) database. They are listed in columns from 13 to 17, respectively. These properties have been extensively described in \cite{2003MNRAS.341...33K,2003MNRAS.341...54K,2003MNRAS.346.1055K}. To ensure clarity, we provide a brief introduction to these properties as follows:

\begin{description}
\item \textbf{4000-\AA break (Dn4000)} is an approximation used to estimate the age of the stellar population in a galaxy (\citealt{1983ApJ...273..105B}). Larger values of Dn4000 indicate galaxies with older stellar populations. In this paper, we measure the Dn4000 value using the narrow version of the index as defined in \cite{1999ApJ...527...54B}.
\item \textbf{Concentration index ($C_{\rm r}$)} is the ratio between the radii that encompass 90\% and 50\% of the light in the r-band, which can be utilized for the purpose of morphological classification of galaxies. Early-type galaxies (ETGs) tend to exhibit higher values of the $C_{\rm r}$ than late-type galaxies.
\item \textbf{r-band absolute magnitude ($M_{\rm r}$)} is defined as the apparent magnitude in the r-band that the object would have if it were viewed at a standard distance of 10 pc (32.6 light-years). 
\item \textbf{Black hole mass ($M_{\rm BH}$)} is derived from the well-established $M-\sigma$ relation, as defined in \cite{2002ApJ...574..740T}. This relation demonstrates a robust correlation between the mass of the central black hole ($M_{\rm BH}$) and the stellar velocity dispersion ($\sigma_\star$).
\item \textbf{[O III] luminosity ($L_{\rm [O\,III]}$)} is calculated from the [O III] 5007 emission line.
\end{description}

Finally, the HERG/LERG classification results are listed in column 18, which used the classification scheme presented in \citealt{2012MNRAS.421.1569B}.

\subsection{Radio properties}

\begin{figure}[!ht]
   \centering
  \includegraphics[width=12cm, angle=0]{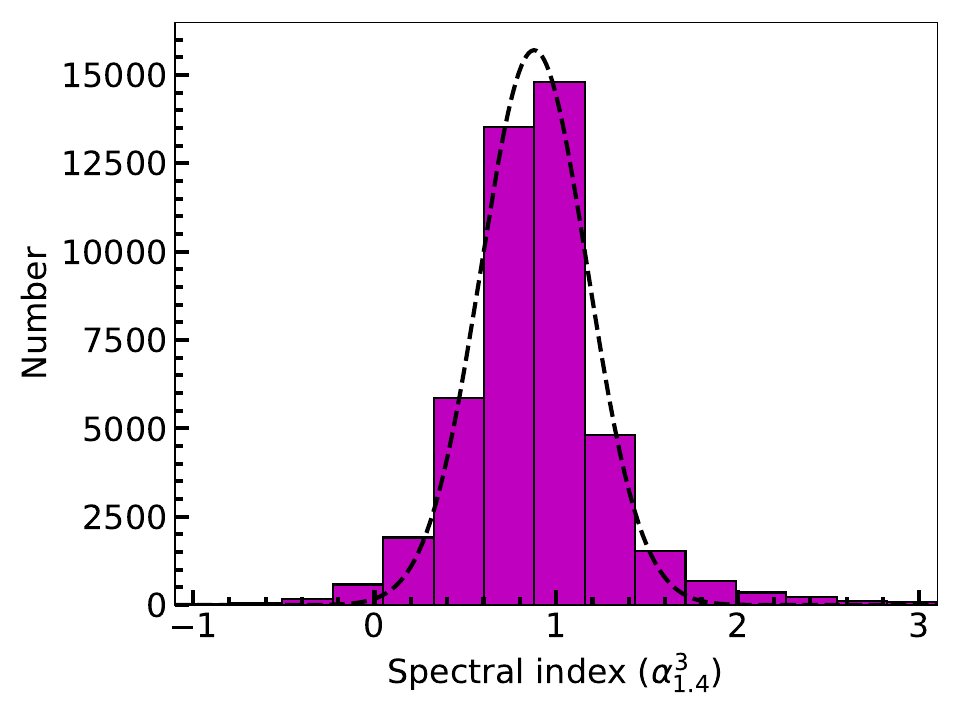}
   \caption{Histogram of the spectral index ($\alpha_{1.4}^3$) distribution for FR-II candidates, showing a mean and median of 0.89 and 0.88, respectively. The black dotted curve shows Gaussian fits to the histogram.} 
   \label{fig:spix}
\end{figure}

We estimate the power-law ($S \propto {v^{-\alpha }}$) radio spectral index between 1.4 and 3\,GHz ($\alpha_{1.4}^3$), where $S$ is the integrated flux density at a specific frequency $v$, and $\alpha$ is the spectral index. The radio spectral index distribution for all the FR-IIs is shown in Figure~\ref{fig:spix}. FR-II sample in this work has a spectral index distribution peaked at 0.89, which is steeper than the typical value of 0.7 for radio galaxies (e.g. \citealt{2010MNRAS.408.2261K}). Most of the FR-II radio galaxies (87.0\%) show a steep radio spectrum with a spectral index greater than 0.5, indicating that the measured spectral index is in agreement with the theoretical value. The 11.4\% of FR-IIs exhibit spectral index ranging between 0 and 0.5, which can be attributed to their complex structures. The remaining FR-IIs (1.6\%) are inverted spectrum and GHz peaked spectrum (GPS) radio galaxies, which have small sizes with an average LAS of 23$^{\prime\prime}$.

We measured the radio luminosity for all FR-II candidates (when both spectral index ($\alpha$) and redshift ($z$) are available) using the following standard formula in \citealt{2009MNRAS.392..617D}:
\begin{equation}\label{eq-L}
L_{\rm rad} = 4\pi D_L^2\left( z \right){S_{1.4}}{\left( {1 + z} \right)^{\alpha_{1.4}^3  - 1}},
\end{equation}
where $D_L$ is the luminosity distance to the source in meter (m), $z$ is the redshift of the radio galaxy, 
$S_{1.4}$ is the measured flux density (${\rm W}\,{\rm m}^{-2}\,{\rm Hz}^{-1}$) at 1.4 GHz, ${\left( {1 + z} \right)^{\alpha  - 1}}$ is the standard $k$-correction used in radio astronomy. 

\begin{figure}[!ht]
   \centering
  \includegraphics[width=12cm, angle=0]{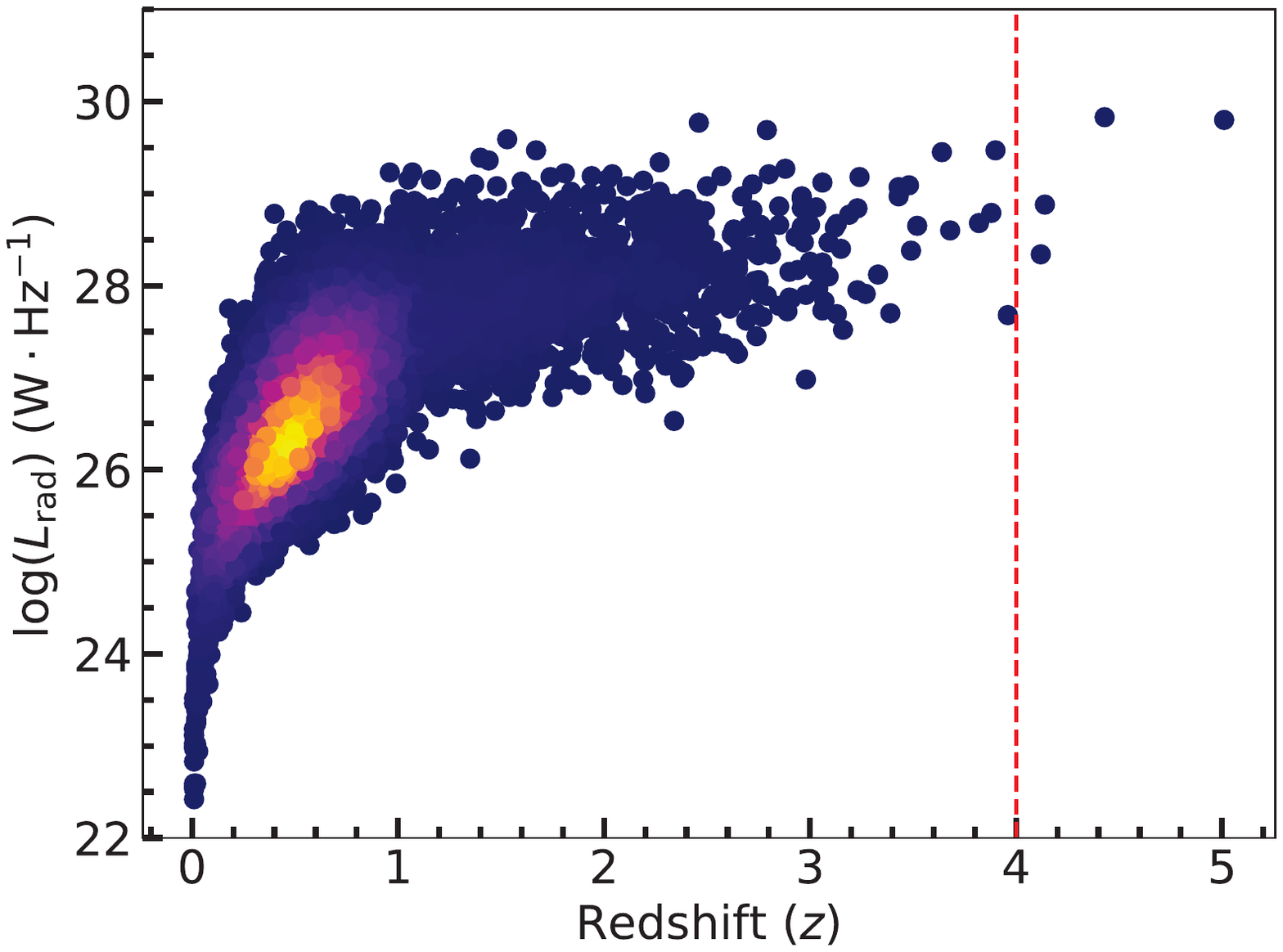}
   \caption{The radio luminosity ($L_{\rm rad}$) at 1.4 GHz distribution of FR-II candidates with redshifts ($z$). The color of each circle in the distribution represents the candidate number density with corresponding redshift and radio luminosity at 1.4 GHz. The red dashed line refers $z=4$.} 
   \label{fig:Lum}
\end{figure}

Figure~\ref{fig:Lum} displays the 1.4 GHz luminosity versus redshift for our FR-II candidates with known redshifts. The mean and median values of radio luminosity for FR-II candidates with available redshifts are found to be log($L_{\rm rad}/{\rm W}\,{\rm Hz}^{-1})=26.61$ and log($L_{\rm rad}/{\rm W}\,{\rm Hz}^{-1})=26.57$, respectively. This is in agreement with the division between the FR-Is and FR-IIs (\citealt{fanaroff1974morphology}). The most radio luminous FR-II candidate is J1239+1005, with $L_{\rm rad}=6.76\times10^{29}\,{\rm W}\,{\rm Hz}^{-1}$. Similarly, the
least radio luminous FR-II candidate is J1109+1206, with $L_{\rm rad}=2.63\times10^{22}\,{\rm W}\,{\rm Hz}^{-1}$. 

\subsection{Host properties}
\begin{figure}[!ht]
   \centering
  \includegraphics[width=7cm, angle=0]{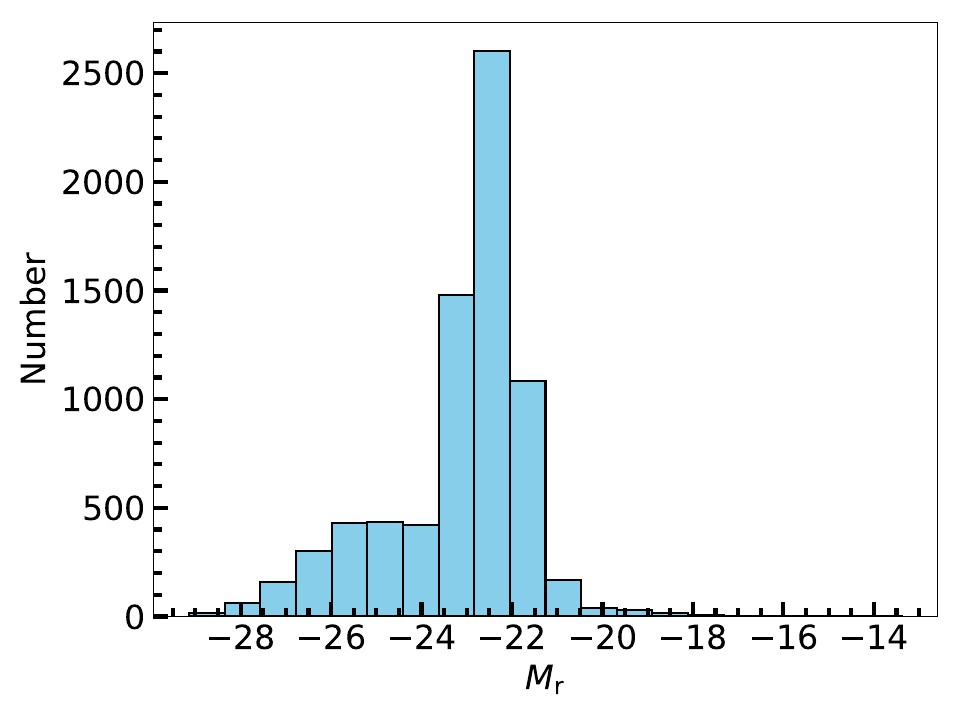}
  \includegraphics[width=7cm, angle=0]{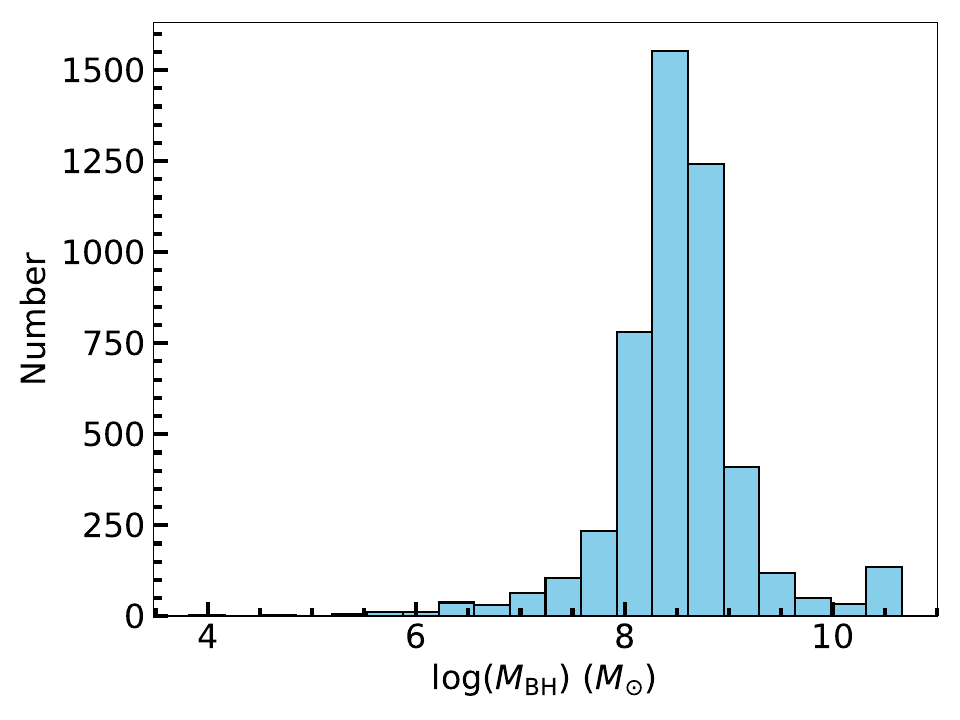}
   \caption{Distributions of the r-band absolute magnitude $M_{\rm r}$ (left panel) and black hole mass $M_{\rm BH}$ (right panel).} 
   \label{fig:Mr_MBH}
\end{figure}

%\subsection{High/low excitation classification of the samples}
The $M_{\rm r}$ were successfully measured for 7,249 FR-II candidates. The distribution of the $M_{\rm r}$ of these candidates is displayed in Figure~\ref{fig:Mr_MBH}, on the left panel, indicating that the hosts cover a range of $-14 \gtrsim M_{\rm r} \gtrsim -29$. Among them, most of the candidates (91.8\%) have $-20 \gtrsim M_{\rm r} \gtrsim -26$. The $M_{\rm BH}$ were successfully measured for 4,837 FR-II candidates. The distribution of $M_{\rm BH}$ shows in Figure~\ref{fig:Mr_MBH}, on the right panel. The majority (92.0\%) of them have $9.5 \gtrsim {\rm log}(M_{\rm BH}) \gtrsim 7.5$ $M_{\odot}$. 
%The J0027+0301 

\begin{figure}[!ht]
   \centering
  \includegraphics[width=7cm, angle=0]{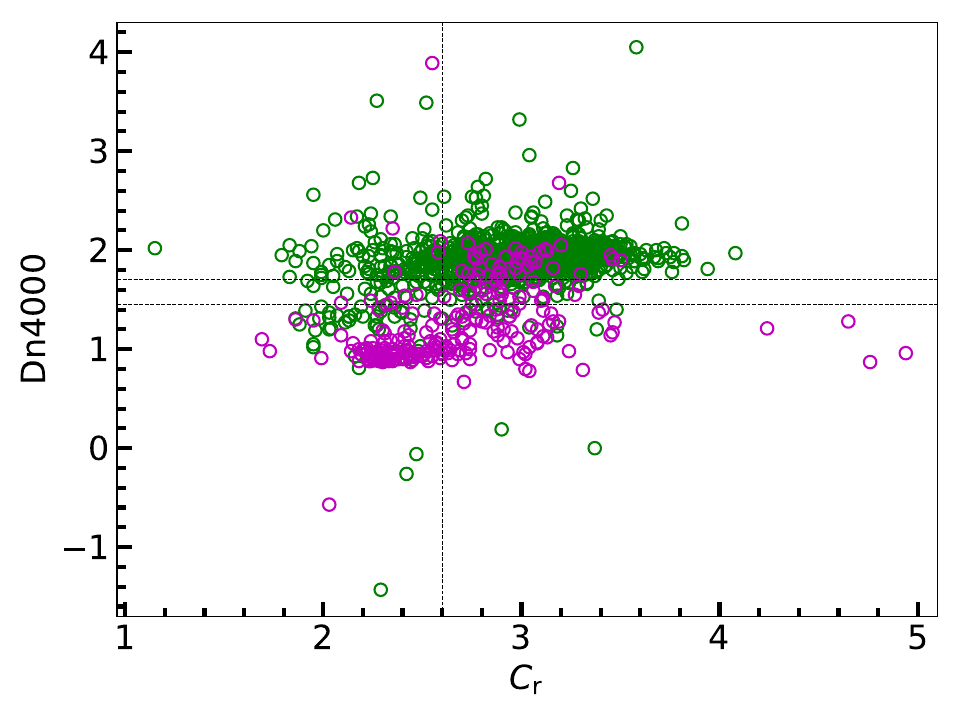}
  \includegraphics[width=7cm, angle=0]{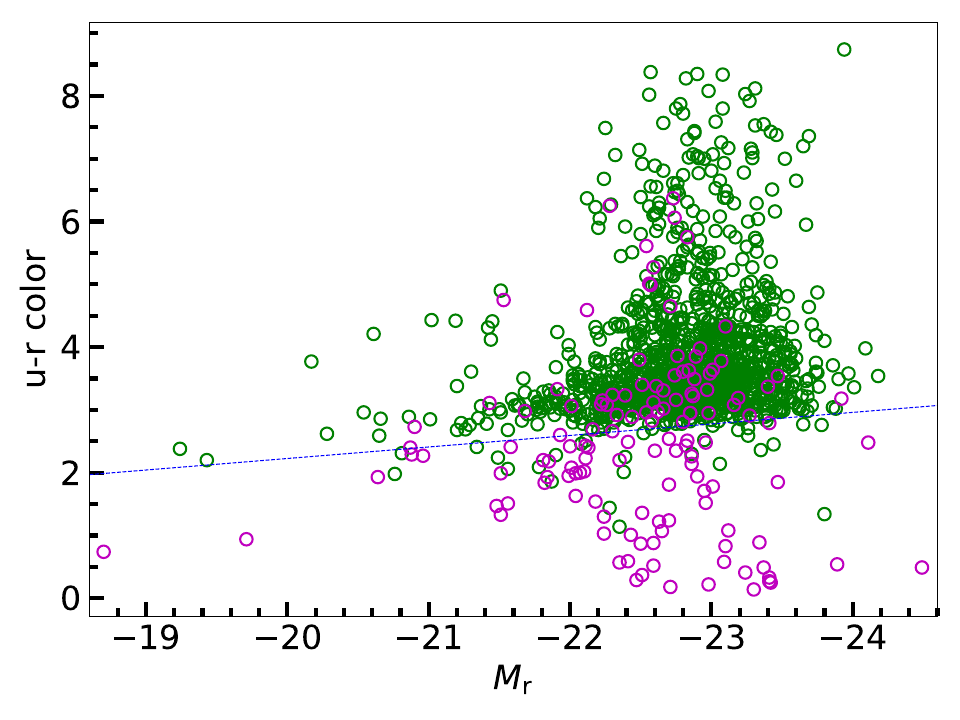}
   \caption{Concentration index $C_{\rm r}$ versus Dn4000 index for the FR-II candidates (left panel). The absolute r-band magnitude ($M_{\rm r}$) versus u-r color for ETG FR-II (right panel). The LERG and HERG FR-II are represented by green and magenta circles, respectively. The blue dashed line in the right panel represents the threshold that separates red and blue ETGs.} 
   \label{fig:Cr_Dn4000}
\end{figure}

The majority (96.3\%) of the FR-II candidates cannot be classified spectroscopically due to the absence of emission lines. Among the remaining candidates, 1,431 FR-IIs are classified as LERG, while 260 FR-IIs are identified as HERG. Galaxies with Dn4000 $\geq1.7$ are defined as red galaxies and galaxies with Dn4000 $\leq1.45$ are defined as blue galaxies (\citealt{2012A&A...541A..62J}). ETGs are defined as galaxies with a more lenient selection criteria at $C_{\rm r} \gtrsim 2.6$ (\citealt{2010MNRAS.404.2087B}). Figure~\ref{fig:Cr_Dn4000} shows the $C_{\rm r}$ versus the Dn4000 index (left panel) for the LERG and HERG FR-II candidates. More than 82\% of the LERG FR-II hosts are located in the region characterized by high values of both $C_{\rm r}$ and Dn4000, indicating that they are red ETGs. On the other hand, more than 71\% HERG FR-II exhibit lower values of Dn4000, suggesting that they are blue galaxies. Within this subset, there are 72 blue ETGs.

Furthermore, the u-r color of the ETG FR-II has been considered.
Figure~\ref{fig:Cr_Dn4000} (right panel) shows the $M_{\rm r}$ of the hosts versus the u-r color. The majority of ETGs are found to lie on a tightly clustered red sequence, while a smaller fraction exhibits a significant scatter towards bluer u-r colors. We applied the method described in \cite{2009MNRAS.396..818S} to fit a Gaussian distribution to the u-r color of the early-type population. With the fitted Gaussian, we determined the 3$\sigma$ offset to the red sequence. Any ETGs below this offset, indicated by the blue dashed line, are defined as blue ETGs. Consistent with previous findings based on the Dn4000 index, the FR-II HERGs display a bluer color compared to the FR-II LERGs. Among the FR-II LERGs, a total of 28 galaxies are found to have an u-r color below the dashed line.

\begin{figure}[!ht]
   \centering
  \includegraphics[width=12cm, angle=0]{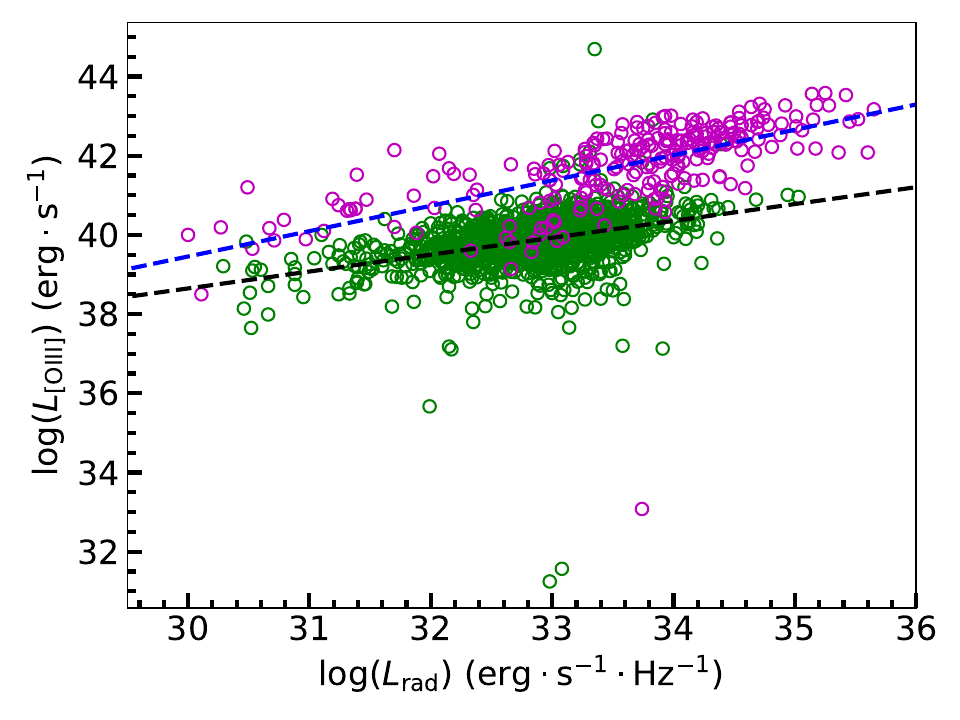}
   \caption{Radio luminosity at 1.4 GHz versus [O III] emission line luminosity of the LERG (green circle) and HERG (magenta circle) FR-II candidates. The dashed blue line represents the linear fit on HERG FR-II candidates. The black dashed line is the linear fit for LERG FR-IIs with a slope of 0.43. The blue dashed line is the linear fit for HERG FR-IIs with a slope of 0.64.} 
   \label{fig:Lrad_Loiii}
\end{figure}

As shown in Figure~\ref{fig:Lrad_Loiii}, LERG and HERG FR-IIs exhibit a broad range of luminosities in both radio and [O III] line emissions, with a significant concentration around ${\rm log}(L_{\rm [O\,III]}/{\rm erg} \cdot {\rm s}^{-1})=40$. 
A correlation can be found in LERG FR-IIs. The linear fit derived from LERG FR-IIs reveals a shallower slope of 0.43, which is less than twice the slope of 0.99 reported for 3C LERGs in \cite{2010A&A...509A...6B}.
Similarly, a correlation can be observed in HERG FR-IIs. The linear fit obtained from HERG FR-IIs indicates a slope of 0.64, which is lower than the slope of 1.15 reported for 3C HERGs in \cite{2010A&A...509A...6B}.

\subsection{GRGs in the sample}
\begin{figure}[!ht]
   \centering
  \includegraphics[width=12cm, angle=0]{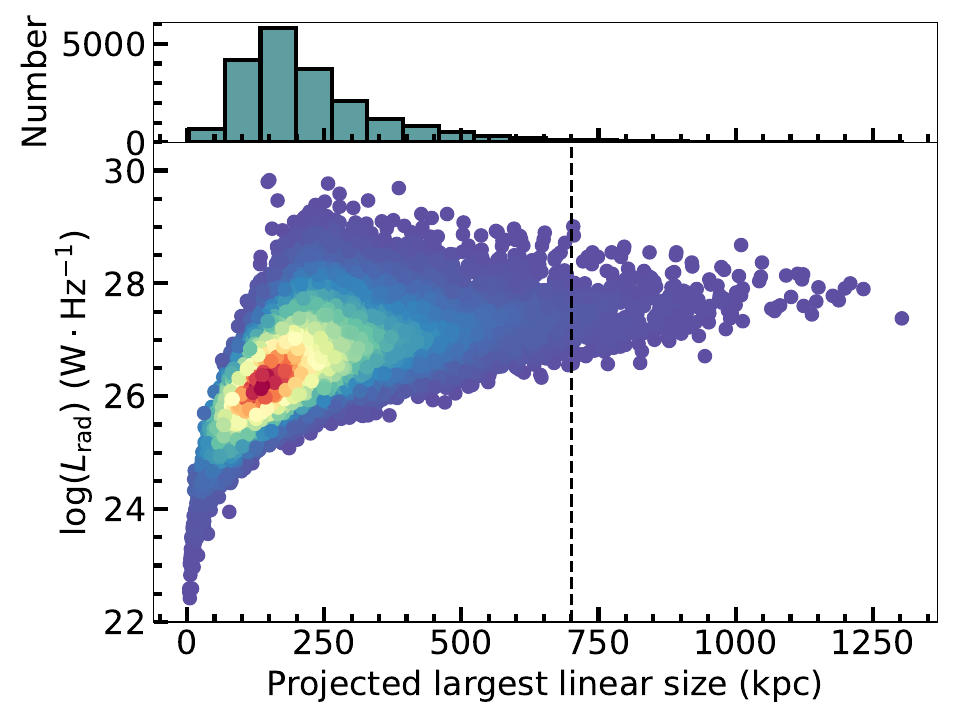}
   \caption{Upper panel: The projected largest linear size (LSS) histogram of the FR-II candidates. Lower panel: The radio luminosity ($L_{\rm rad}$) at 1.4 GHz distribution of FR-II candidates LSS. The color of each circle in the distribution represents the candidate number density with corresponding LSS and $L_{\rm rad}$. The black dashed line represents 700 kpc of LLS.} 
   \label{fig:LSS}
\end{figure}

The angular size distribution of the FR-II sample FRIIRGcat is from 9$^{\prime\prime}$ to 214$^{\prime\prime}$. For those with redshift determined, we can estimate their physical sizes. In total, the physical sizes (the largest linear size, LLS) of 20,147 FRIIs are estimated, ranging from 4 to 1,209\,kpc. Figure~\ref{fig:LSS} shows the LLS-$L_{\rm rad}$ distribution of our FR-II candidates. The majority of the candidates share modest radio luminosity and LLS. FR-IIs with larger LLS tend to have higher radio luminosity.

\begin{figure}[!ht]
   \centering
   \includegraphics[scale=0.18]{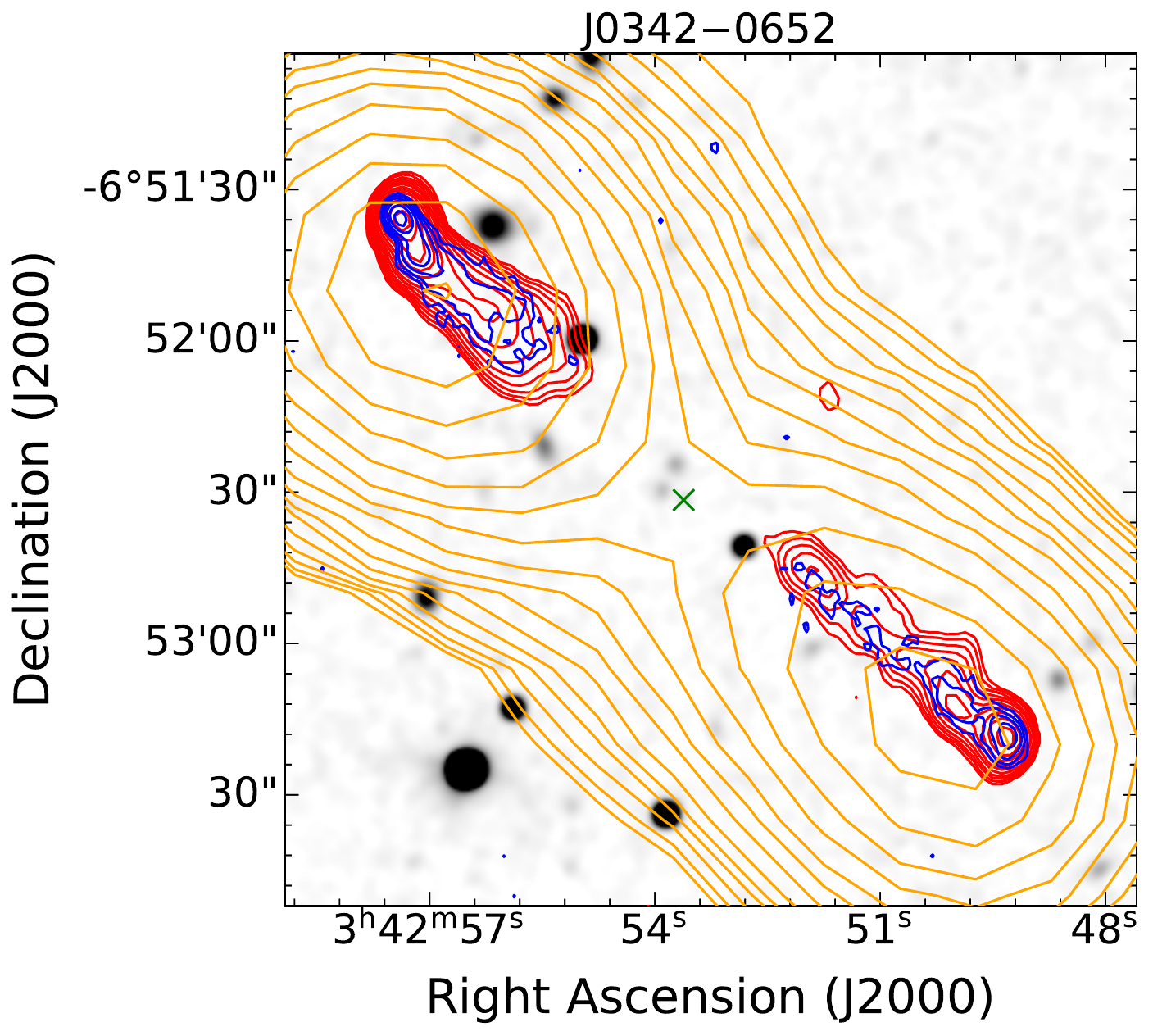}
   % \hspace{1mm}
   % \vspace{0.2mm}
   \includegraphics[scale=0.18]{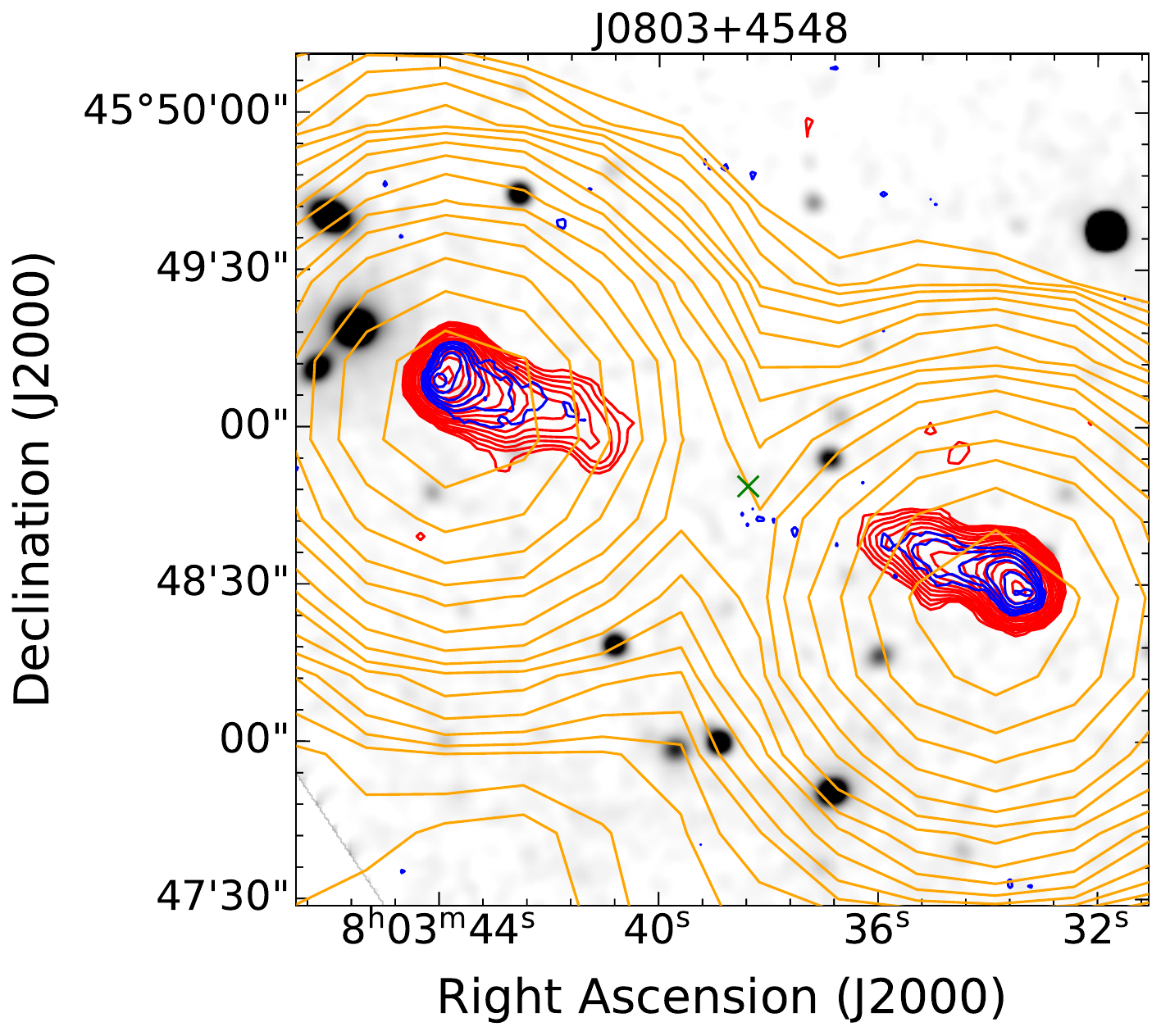}
   \includegraphics[scale=0.18]{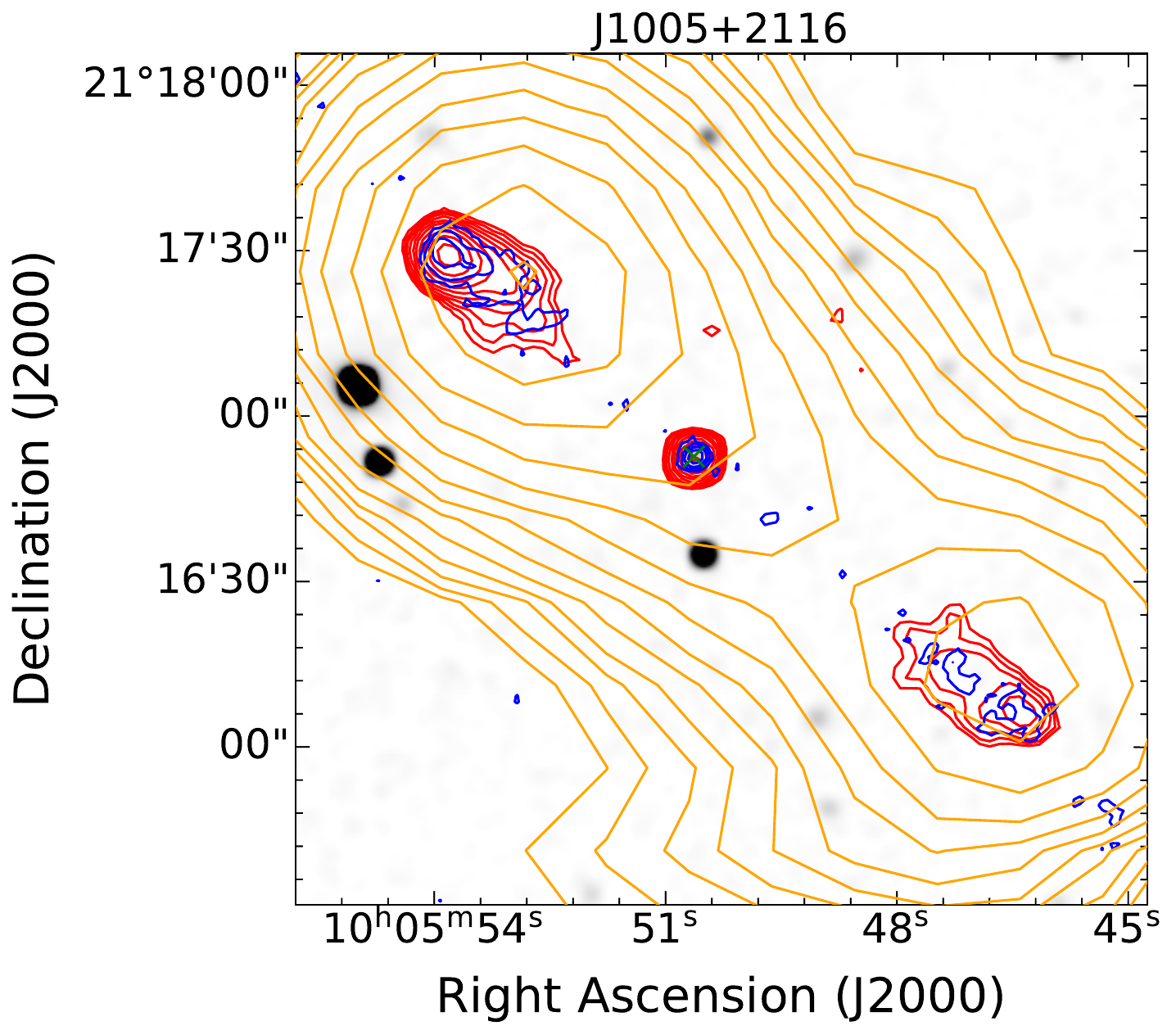}
   % \hspace{1mm}
   \includegraphics[scale=0.18]{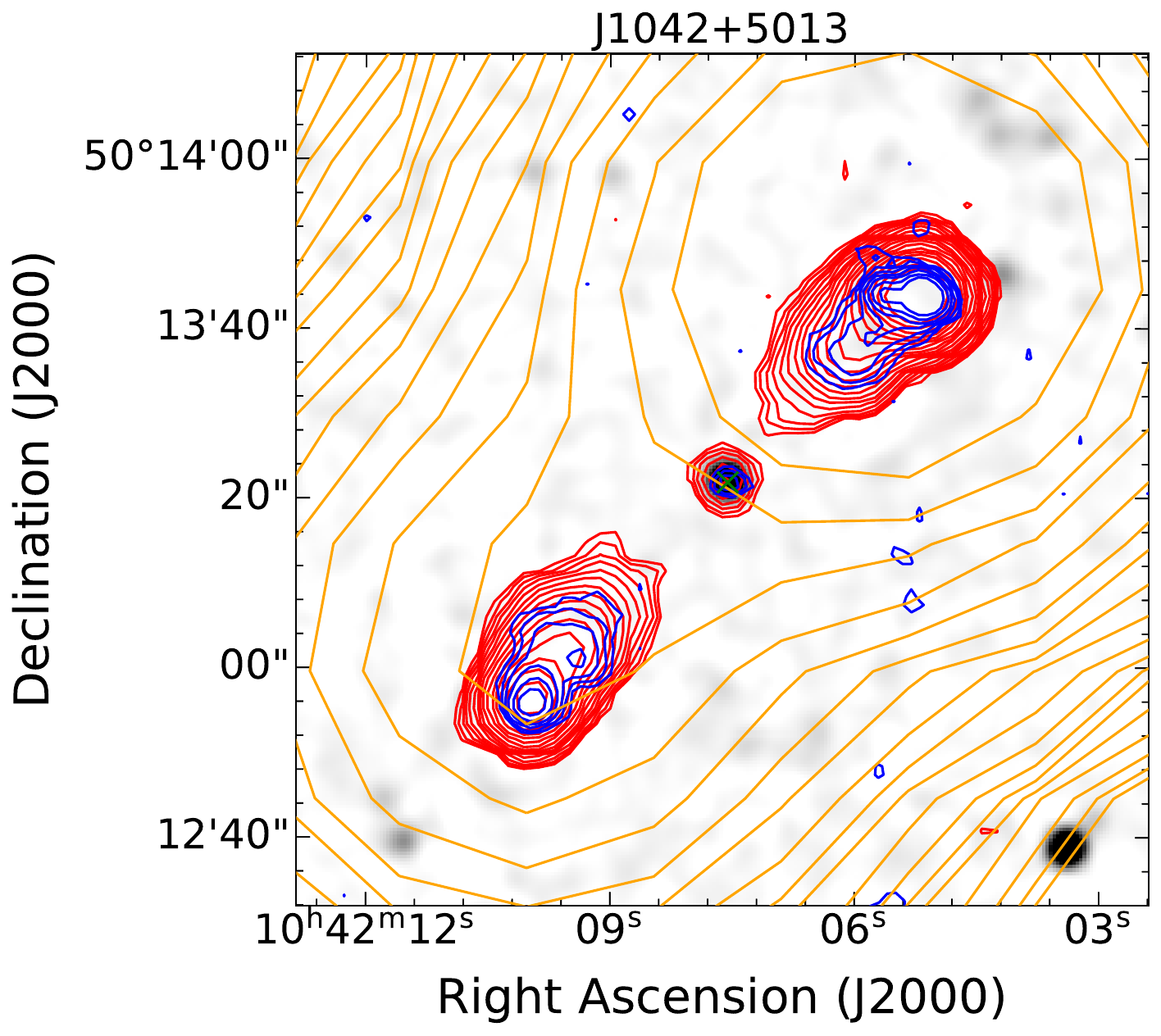}
   % \hspace{1mm}
   \includegraphics[scale=0.18]{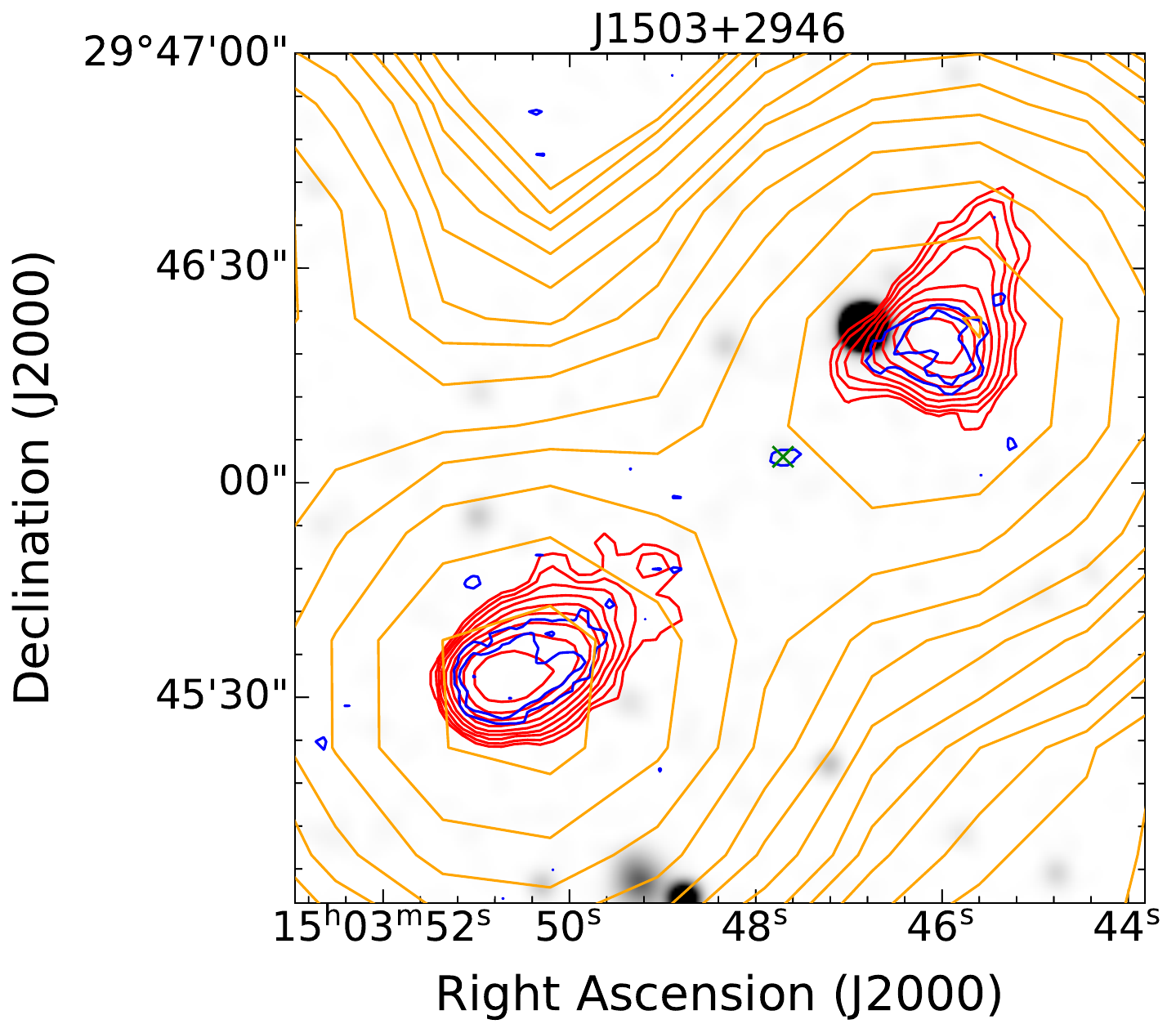}
   \includegraphics[scale=0.18]{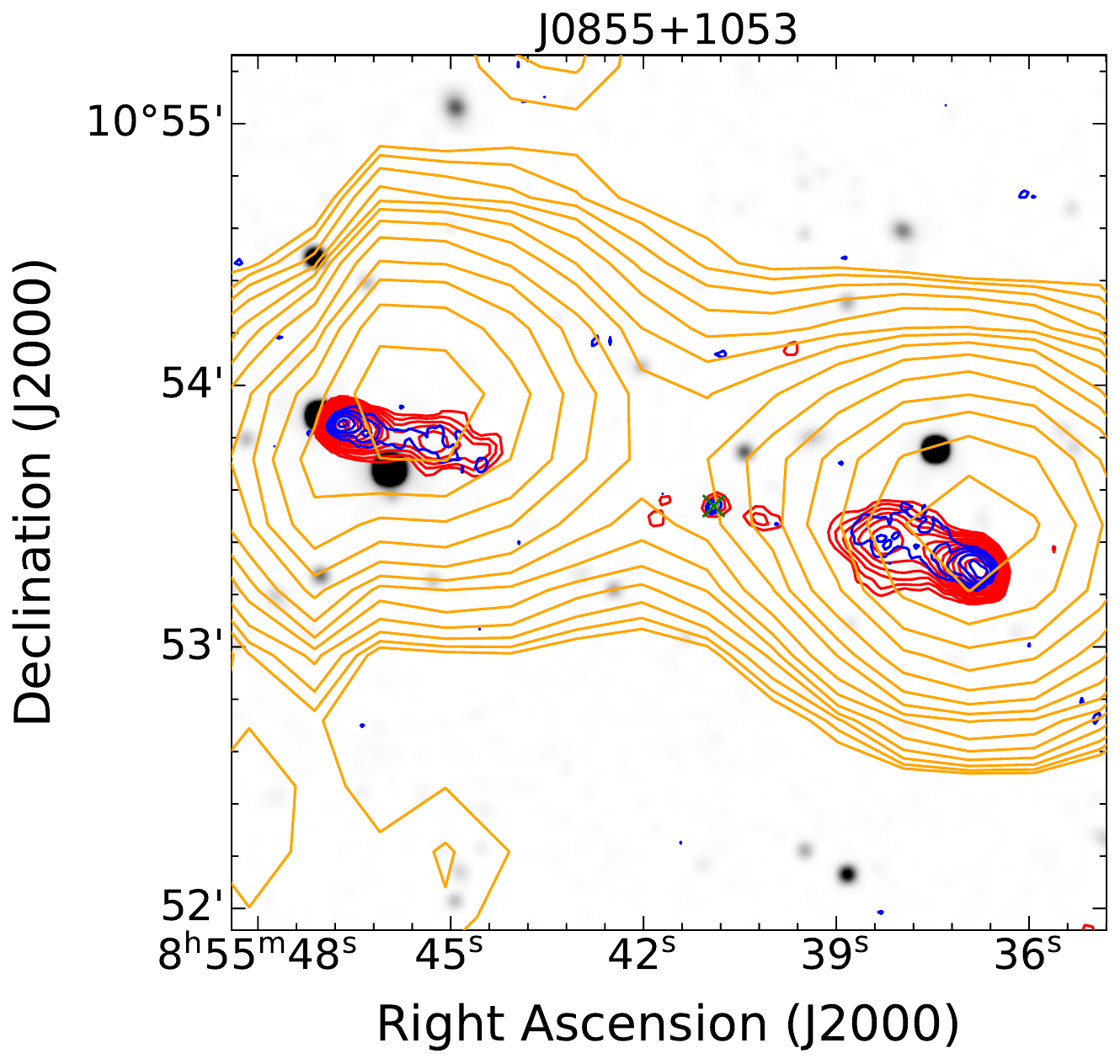}
   \includegraphics[scale=0.18]{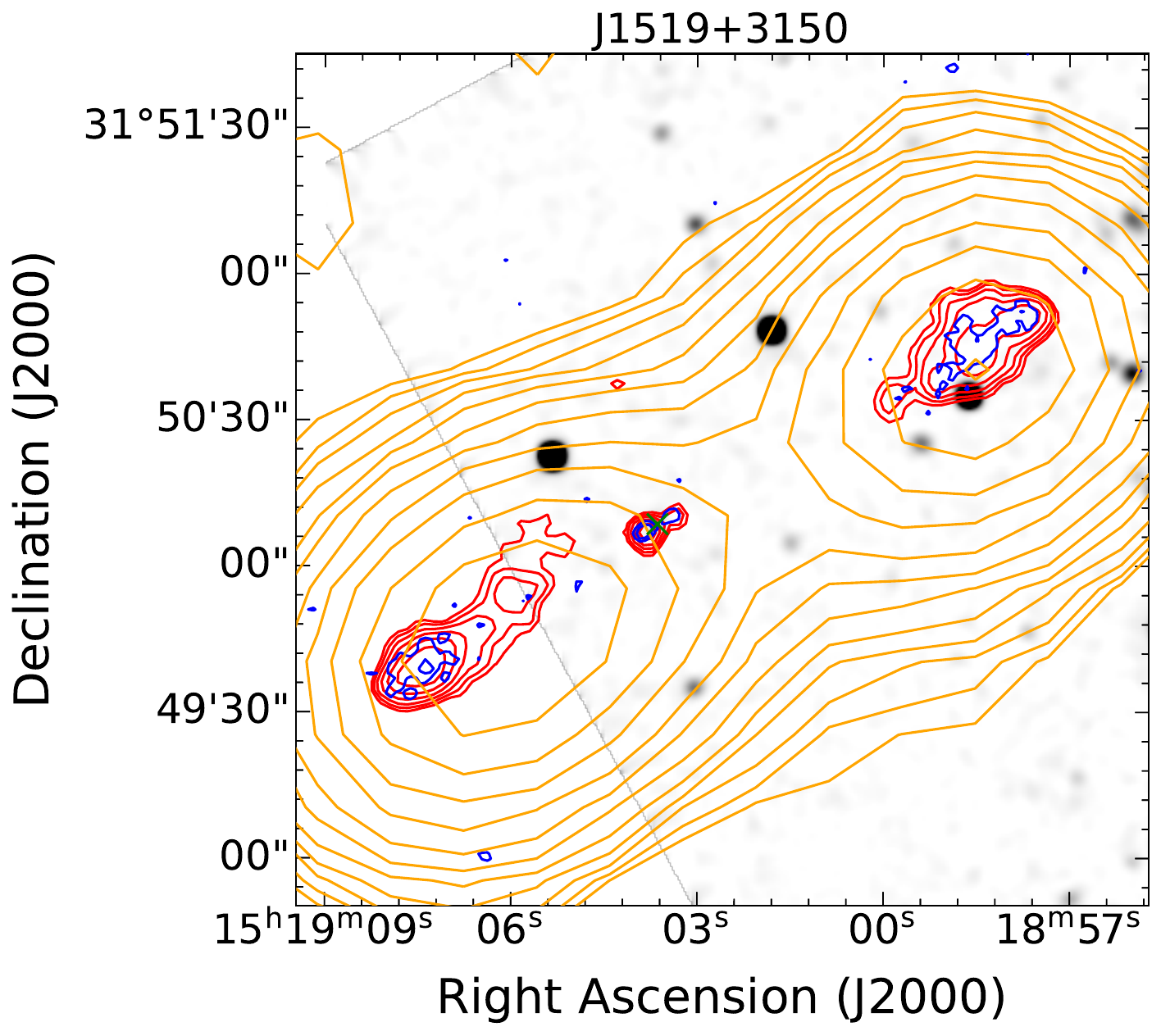}
   % \hspace{1mm}
   \includegraphics[scale=0.18]{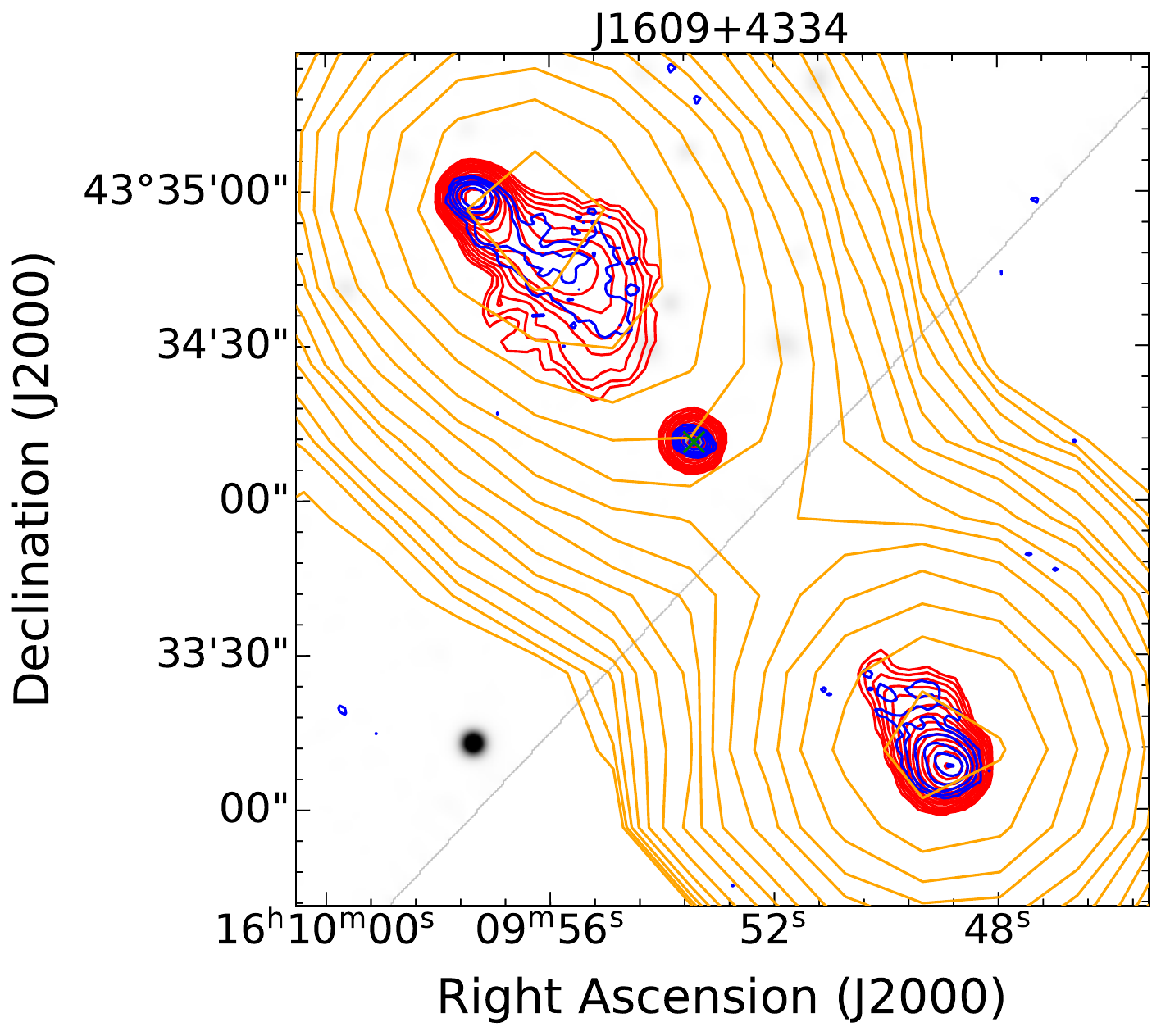}
   \includegraphics[scale=0.18]{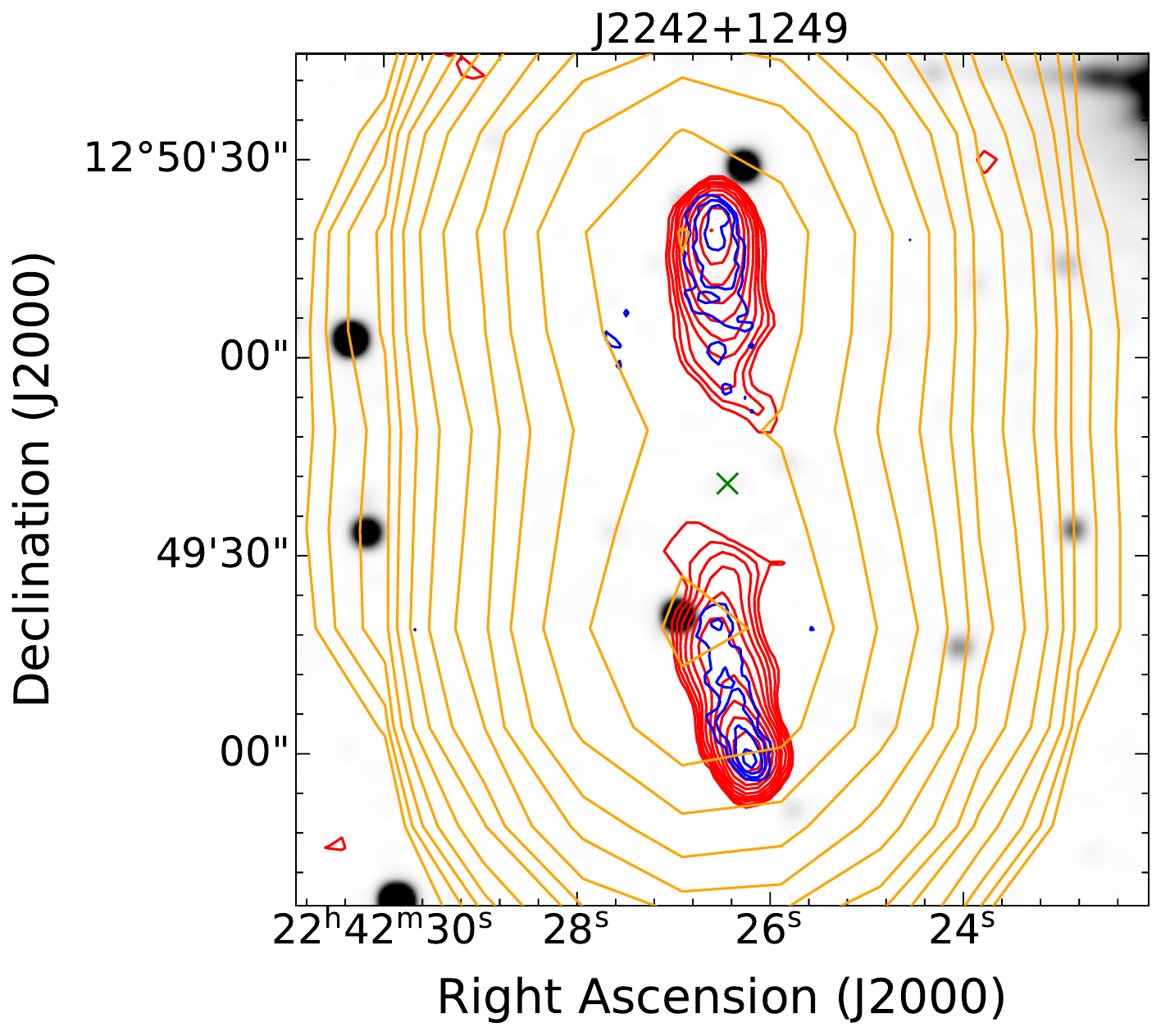}
   % \hspace{1mm}
   \caption{Radio and optical maps of a sample of 9 GRGs. The contour lines of red, blue, and orange colors represent the radio emissions detected in the FIRST, VLASS, and NVSS images correspondingly. The background images are from the SDSS r-band. Contour levels are at $3\sigma\times$[1, 1.41, 2, 2.83, 4, 5.66, 8, 11.31, 16, 22.63, 32, 45.25, 64, 90.51, 128, 181.02, 256, 362.04], where $\sigma$ is the local rms noise. The green cross is centered on the host galaxy.} 
\label{fig:GRGs_radio}
\end{figure}

GRGs are now generally defined as those with projected linear sizes larger than $700$\,kpc, which ensues the classification of 307 GRGs in our FR-II-type radio galaxy sample. We note that a few of them has already listed in the existing GRG samples \cite{2018ApJS..238....9K}, \cite{2020A&A...635A...5D, 2020A&A...642A.153D}, \cite{2020MNRAS.499...68T}, and \cite{2020ApJS..247...53K}. We over-viewing the 307 GRG sample from the literature and result in 5 (J0857+3945, J1030+5310, J1236+1034, J1511+0751, and J2356-0131), 12 (J1057+5105, J1236+4604, J1256+5528, J1312+4809, J1334+5501, J1336+5540, J1341+4916, J1406+5547, J1407+5132, J1419+4837, J1426+5129, and J1504+5030), 3 (J0847+4223, J1129-0121, J1551+1035), and 3 (J0941+3126, J1331+2357, and J1421+1016) already confirmed GRGs from \cite{2018ApJS..238....9K}, \cite{2020A&A...635A...5D}, \cite{2020A&A...642A.153D}, and \cite{2020MNRAS.499...68T}, respectively. The identification of GRGs that already exist in historical samples confirms the success of our method. The remaining 284 GRG sample identified in this work corresponds to the new identification. In Figure~\ref{fig:GRGs_radio}, we illustrate 9 GRGs among the 284 samples, including the largest (J1609+4334) and the smallest one (J1503+2946), they have the largest projected sizes of 1,209 and 701\,kpc, respectively. The new GRG sample has redshift spans $0.31<z<2.42$.

\subsection{Alignment in the orientation of the samples}

For the sake of maintaining consistency, this particular phase of the experiment specifically targeted FR-II candidates in the northern sky region, due to observations for half of the southern sky region were conducted after transitioning to the new Expanded Very Large Array (EVLA) (\citealt{2015ApJ...801...26H}). After implementing this exclusion criterion, 36,908 FR-II candidates were selected, which covers a sky area of $\sim$8,444 ${\rm deg}^2$, resulting in a number density $\sim 4$ ${\rm deg}^{-2}$.

\begin{figure}[!ht]
   \centering
  \includegraphics[width=12cm, angle=0]{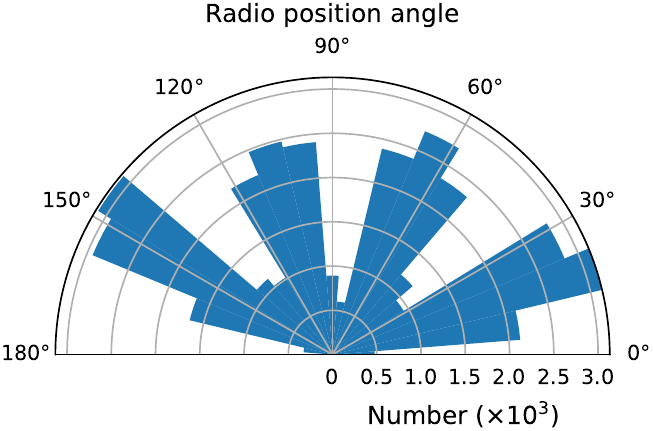}
   \caption{Radio position angle (RPA) distribution of the complete sample of selected FR-II galaxies.} 
   \label{fig:RPAs}
\end{figure}

The distribution of radio position angles for all selected FR-II candidates is given in Figure~\ref{fig:RPAs}. Normally, in the absence of systematic effects, the radio position angles would be uniformly distributed across this large area of the sky. In Figure~\ref{fig:RPAs}, the distribution exhibits four prominent peaks around 25, 65, 115, and 150 degrees, suggesting the presence of systematic effects in these directions. Observing beams effect in the test data comprise the dominant aspect of systematic effects. Specifically, the sidelobes in the beam patterns surrounding the bright sources were distributed in these four directions (\citealt{2015ApJ...801...26H}). Similar effects were observed and discussed in \cite{2017MNRAS.472..636C}, which have been proved to be non-local, thus having no impact on our analysis.

To further investigate the conformity of the distribution with a uniform distribution of the chosen galaxies, the Kolmogorov-Smirnov (K-S) test (e.g., \citealt{2012JCAP...01..009F}) was utilized for analysis. The K-S test yielded a $p$-value close to zero percent, providing substantial evidence for rejecting the null hypothesis that the radio position angle distribution across the entire selected galaxy sample is uniform. 

\begin{figure}[!ht]
   \centering
  \includegraphics[width=12cm, angle=0]{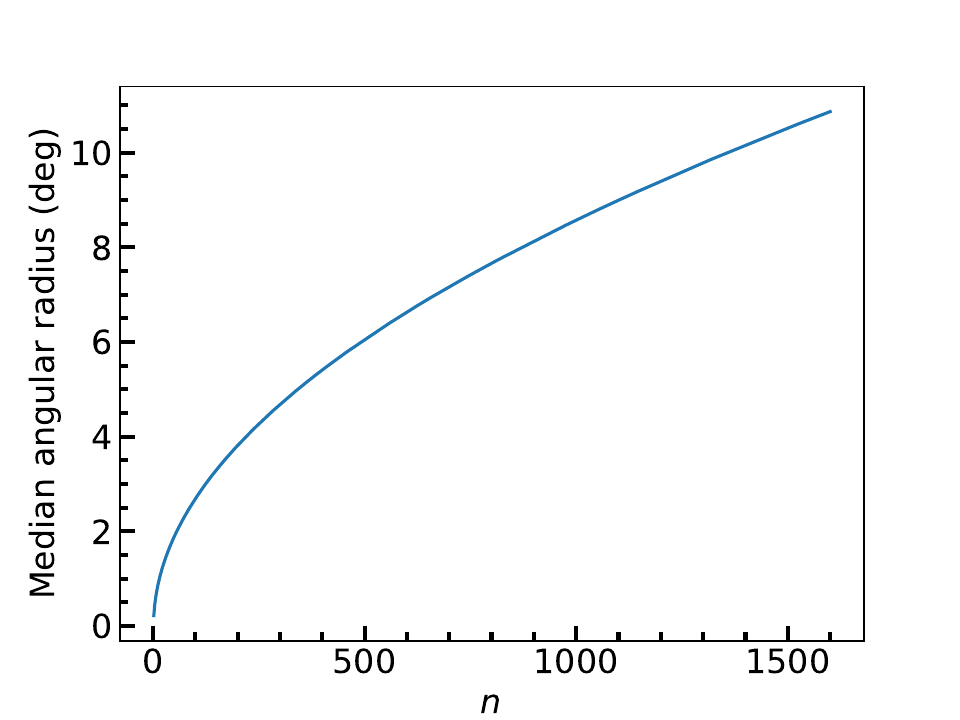}
   \caption{
Illustration of the relationship between the median angular radius and the number of nearest neighbors $n$. The median angular radius is obtained by encircling each galaxy with a circular aperture, where the size of the aperture is determined by the distance to the $n$th neighboring galaxy.
} 
   \label{fig:angle_vs_n}
\end{figure}

To estimate the uncertainties and determine the significance level in rejecting the null hypothesis on different angular scales, we compared the results of the selected FR-II candidates with those of 1,000 simulated position angle samples that were randomly distributed. These samples were generated by shuffling the position angles among the galaxies while maintaining the overall position angle distribution and the positions of the galaxies. To represent the statistics in terms of angular scale, a circular aperture is plotted with a radius extending to the $n$th neighbor of each galaxy. The number of nearest neighbors is then converted to the approximate corresponding angular scale by calculating the median angular radius of all these apertures. This relationship is illustrated in Figure~\ref{fig:angle_vs_n}.

\begin{figure}[!ht]
   \centering
  \includegraphics[width=12cm, angle=0]{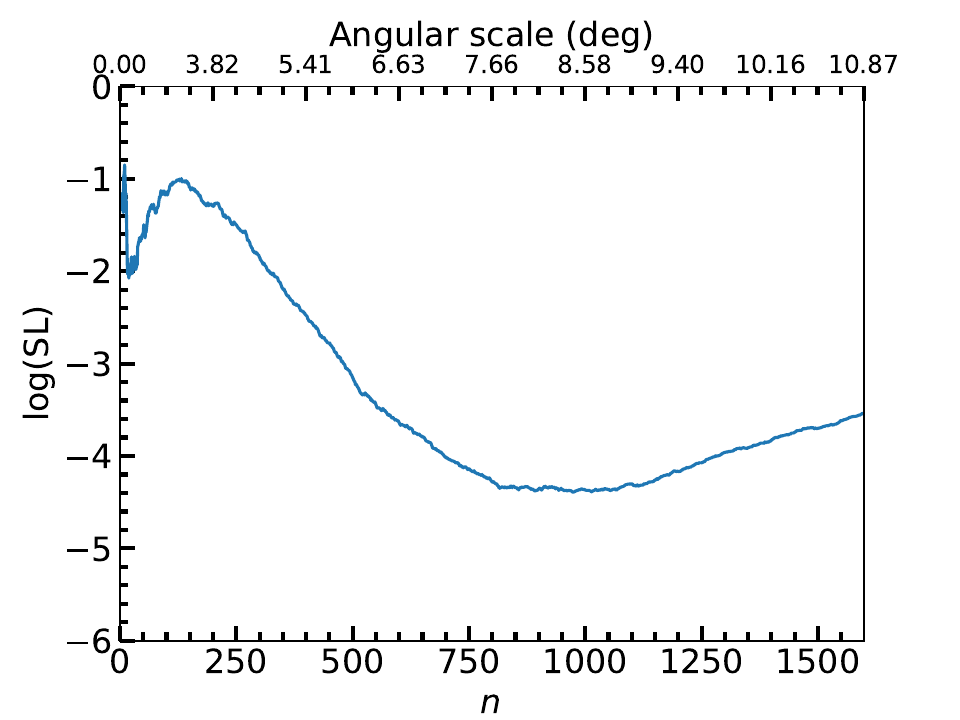}
   \caption{The distribution of the logarithm of the significance level (SL) as a function of the number of nearest neighbors ($n$) or the corresponding angular scale is shown for the entire sample of selected FR-II candidates. The conversion to angular scale is illustrated in Figure~\ref{fig:angle_vs_n}.
} 
   \label{fig:SL_vs_n}
\end{figure}

\begin{figure}[!ht]
   \centering
  \includegraphics[width=8cm, angle=90]{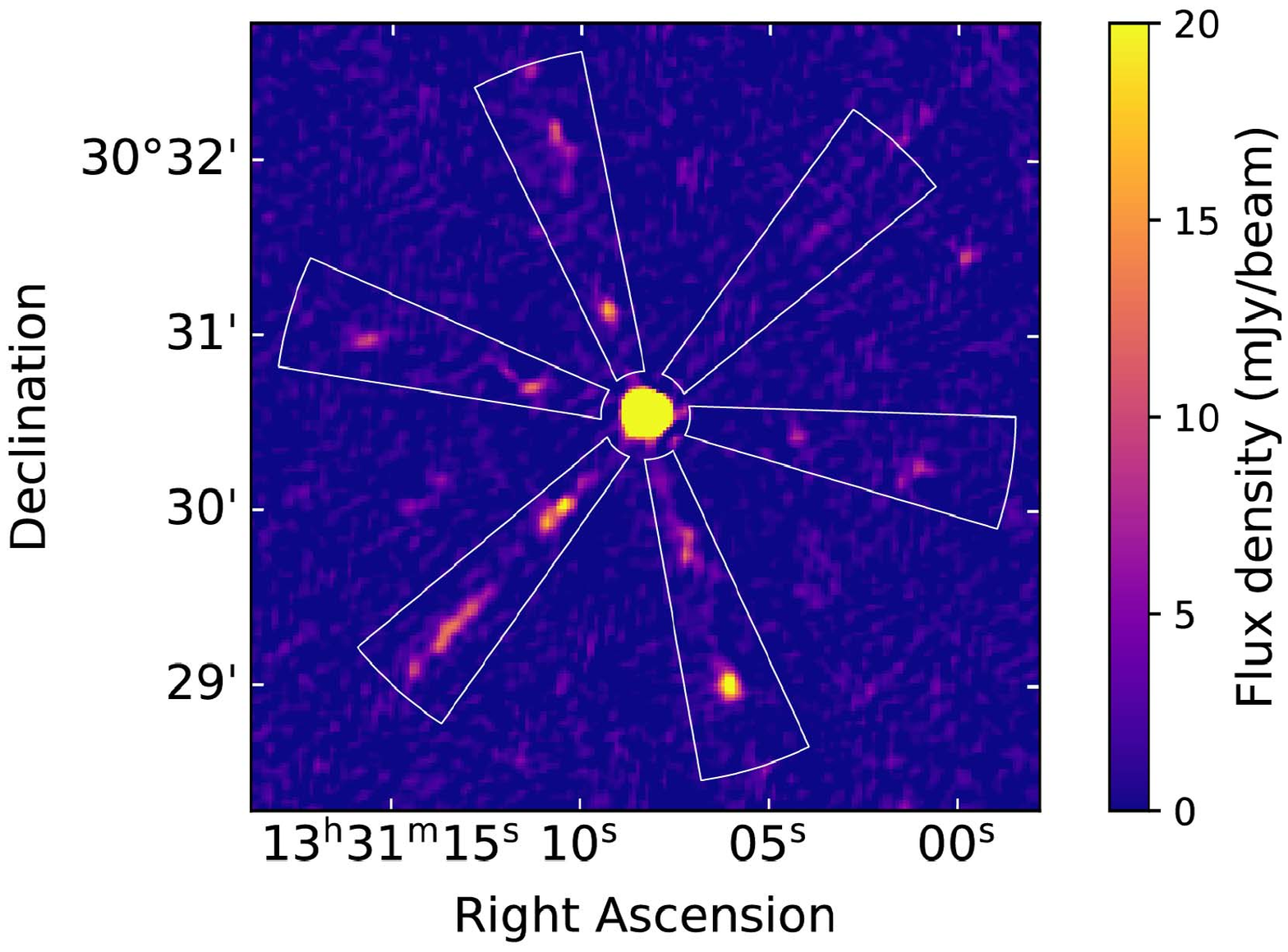}
   \caption{An example to highlight the sidelobe residuals enveloping a bright source, where the main sidelobes are indicated in white regions.} 
   \label{fig:beam_effects}
\end{figure}

The significance level at which the null hypothesis of radio position angles being uniformly distributed should be rejected is presented as a function of the number of nearest neighbors $n$ (or corresponding angular scale) in Figure~\ref{fig:SL_vs_n}. There is strong evidence to reject the hypothesis of uniformity in radio position angles (for alignment) at angular scales of approximately 8 deg, with a significance level of less than $10^{-4}$. As evident from the aforementioned RPA analysis, this alignment effect is most likely caused by systemic effects (\citealt{2017MNRAS.472..636C,2020A&A...642A..70O}). Here, we present an example of the beam pattern in the FIRST image depicted in Figure \ref{fig:beam_effects}, showcasing the sidelobes residuals in a bright source due to the sparse uv-coverage of FIRST snapshot observations. These sidelobes contribute a significant spurious structure to nearby sources and itself (e.g. \citealt{2015ApJ...801...26H}). Therefore, we attribute the observed alignment in our 2D analysis to the sidelobe effect as well.

\begin{figure}[!ht]
   \centering
  \includegraphics[width=12cm, angle=0]{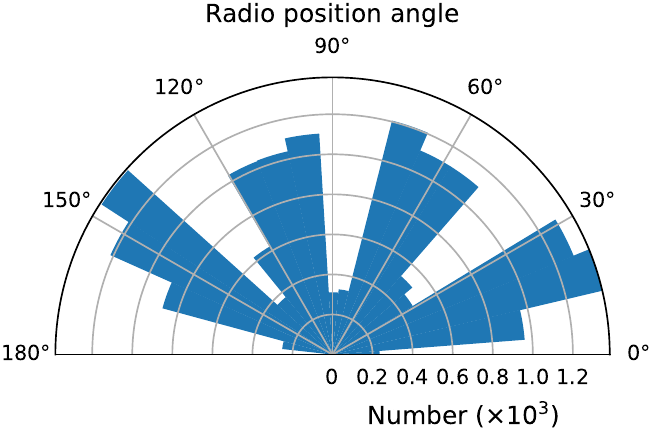}
   \caption{Radio position angle (RPA) distribution for the sample of 16,828 sources that have a redshift measurement.} 
   \label{fig:RPAs_redshift}
\end{figure}

Furthermore, the alignment uniformity using 3D galaxy positions has also been analyzed. The 3D samples were obtained by excluding galaxies lacking redshift information, in the northern sky region (36,908 FR-IIs in total). As a result, the filter reduces the sample size to 16,828 galaxies. The RPA distribution of these samples is illustrated in Figure~\ref{fig:RPAs_redshift}, exhibiting similar systematic effects observed in the complete sample of selected FR-II candidates. The K-S test also yields a $p$-value close to zero percent, suggesting strong evidence to reject the null hypothesis of uniformity of radio position angles for this sample of 16,828 galaxies. Both 2D and 3D analysis have been used for these samples. The statistical method utilized is identical to that employed in the complete sample mentioned above, with the distinction that the 3D analysis is used to determine the $n$ nearest neighbors for each galaxy. 

\begin{figure}[!ht]
   \centering
  \includegraphics[width=12cm, angle=0]{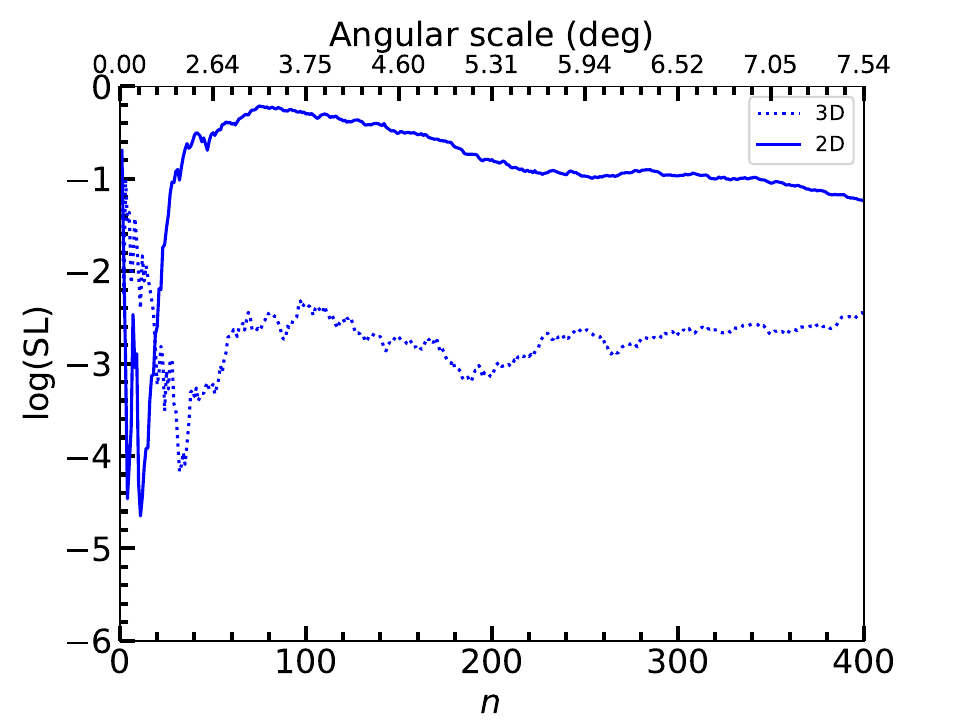}
   \caption{The logarithm of the significance level (SL) at which the rejection of position angle uniformity should occur, as a function of the number of nearest neighbors $n$, is plotted for the 16,826 galaxies with available redshifts. The solid line represents the results of the 2D analysis, while the dashed line represents the results of the 3D analysis.
} 
   \label{fig:SL_vs_n_3D}
\end{figure}

\begin{figure}[!ht]
   \centering
  \includegraphics[width=12cm, angle=0]{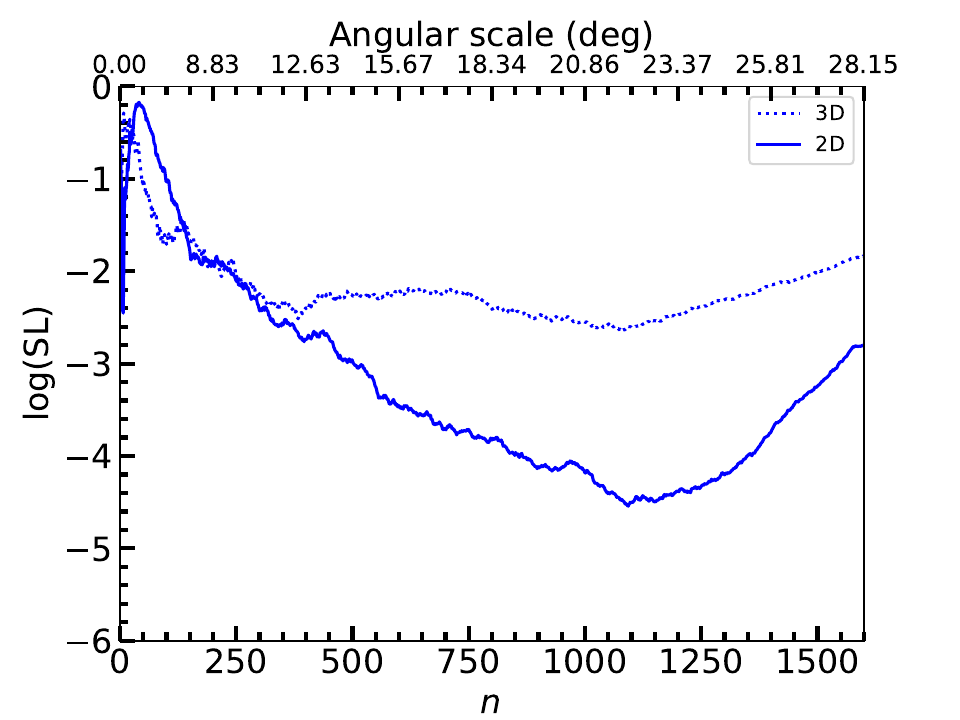}
   \caption{The logarithm of the significance level (SL) at which the rejection of position angle uniformity should occur, as a function of the number of nearest neighbors $n$, is plotted for the 6,132 FR-IIs with available spectroscopic redshifts. The solid line represents the results of the 2D analysis, while the dashed line represents the results of the 3D analysis.
} 
   \label{fig:SL_vs_n_3D_sp}
\end{figure}

Figure~\ref{fig:SL_vs_n_3D} displays the significance level at which the hypothesis of uniformity in position angles can be rejected for the 16,828 galaxies with a redshift. The 2D and 3D analyses indicate substantial evidence of alignment at angular scales of approximately 0.5 and 2.0 degrees, respectively, with a significance level of less than $10^{-4}$. The stronger alignment in the 2D analysis, as indicated by a lower minimum significance level compared to the 3D analysis, suggests that the 2D alignment effect is attributable to the systematic effects (\citealt{2020A&A...642A..70O}). Uncertainties in redshift measurements may cause the 3D alignment effect. We conducted tests to determine if the results depended on whether the redshifts were obtained through photometric or spectroscopic methods. Figure \ref{fig:SL_vs_n_3D_sp} shows the results for the 6,132 FR-IIs that have a spectroscopic redshift. The figure demonstrates that there is no significant evidence of the 3D alignment effect in the analysis of these sub-samples. However, in the 2D analysis, the alignment effect is still observed. This indicates that the presence of the 3D alignment effect in radio galaxies can be influenced by less reliable redshift measurements. This further substantiates the notion that the 2D alignment signal is attributed to systematic effects. These results are consistent with the findings of the study conducted in \cite{2017MNRAS.472..636C} using the same survey data.

Several studies (\citealt{2017MNRAS.472..636C, 2020A&A...642A..70O, 2020MNRAS.499.1226P, 2021A&A...653A.123M}) have reported a 2D alignment, with angular scales of less than 4 deg. These angular scales are similar to those that we have examined in our 3D space samples, whereas half of those we have investigated in our 2D space samples. The differences in the 2D alignment angular scales are due to the utilization of a different and more extensive sample of radio galaxies in this paper. This suggests that the 2D alignment of radio galaxies can occur on larger scales. However, the reports of a 3D alignment (\citealt{2017MNRAS.472..636C, 2020A&A...642A..70O, 2020MNRAS.499.1226P}) have not yielded statistically significant results.

\section{Conclusions}
\label{sect:conclusion}

In this paper, we present FRIIRGcat, a catalog of FR-II radio galaxies consisting of 45,241 candidates from the FIRST survey. The code and results for FRIIRGcat are available on Github\footnote{\url{https://github.com/lao19881213/RGC-Mask-Transfiner/tree/main/tools/inference/FIRST/properties_analysis}}. A total of 41,425 candidates have been identified with optical/infrared counterparts from SDSS DR16 and NED. Among them, 20,417 candidates have measured redshifts ranging from 0.01 to 5.01. The integrated flux densities at 1.4 GHz (NVSS or FIRST) and 3 GHz (VLASS) have been measured for each FR-II. The spectral index between 1.4 GHz and 3 GHz has a mean of 0.89 and a median of 0.88 for all candidates. Most FR-IIs are steep spectrum radio galaxies. The radio luminosity at 1.4 GHz for candidates with available has a mean and median value of log($L_{\rm rad}$)=26.61 ${\rm W}\,{\rm Hz}^{-1}$ and log($L_{\rm rad}$)=26.57 ${\rm W}\,{\rm Hz}^{-1}$. 

Most candidates with available r-band absolute magnitude fall within the range of $-20 \lesssim M_{\rm r} \lesssim -26$. Almost all of the candidates with available black hole mass are found within the range of $7.5 \lesssim {\rm log}(M_{\rm BH}) \lesssim 9.5$. A total of 1,691 FR-IIs have been successfully classified as LERG (1,431) or HERG (260). More than 82\% of LERGs are identified as red ETGs, while more than 71\% of HERGs are identified as blue galaxies. We have discovered a correlation between radio and [O III] line luminosity in our HERGs.  

Based on the definition of GRGs, a total of 284 GRGs have been identified in our FRIIRGcat. These GRGs cover a redshift range of $0.31<z<2.42$ and were previously not included in any existing GRG catalogs. These samples will be one of the important data for the in-depth study of GRGs. Additionally, they were automatically identified utilizing the results from the deep learning detector \HeTu-v2, demonstrating the feasibility of employing deep learning techniques for the fully automatic identification of GRGs.

We analyze the alignment in position angles of 36,908 FR-II candidates at the northern sky area in the FRIIRGcat. The analysis was conducted on the 2D space (36,908) and 3D space (16,828) samples, which were determined based on the absence or presence of redshift information in the original samples. We discovered evidence of alignment in both 2D and 3D space samples. The 2D alignment is most likely due to system effects (e.g. the strong sidelobes in beam patterns), whereas the 3D alignment is likely caused by the less reliable redshift (i.e. photometric redshifts) because there is no alignment signal was observed in the 3D sample with spectroscopic redshifts. The angular scale of the 2D alignment in this work is twice that of the previous works. 
In the case of reliable redshift measurements, this study confirms that the alignment of radio galaxies identified from the FIRST survey is primarily attributed to systematic effects.
This work provides important insights for further investigating the cosmic alignment of AGN jets, which are indicators of large-scale structures and important for cosmological models and parameters. These effects will be important for cosmological analyses involving radio data, such as weak lensing studies conducted with advanced high-resolution radio interferometers like the Square Kilometre Array (SKA; \citealt{2020PASA...37....2W}).

\normalem
\begin{acknowledgements}
%We thank the referee for a constructive report and Jianguo Wang, Cheng Hu and Zhaoyu Chen for great helps on the spectral analysis.
This work was supported by the National SKA Program of China (2022SKA0120101, 2022SKA0130100, 2022SKA0130104), the National Natural Science Foundation of China (Nos. 12103013), the Foundation of Science and Technology of Guizhou Province (Nos. (2021)023), the Foundation of Guizhou Provincial Education Department (Nos. KY(2021)303, KY(2020)003, KY(2023)059). YXL was supported by the National Science Foundation of China (12103076, 12233005), the National Key R\&D Program of China (2020YFE0202100), the Shanghai Sailing Program (21YF1455300), and the China Postdoctoral Science Foundation (2021M693267). 
This research work made use of public data from the Karl G. Jansky Very Large Array (VLA); the VLA facility is operated by the National Radio Astronomy Observatory (NRAO). The NRAO is a facility of the National Science Foundation operated under cooperative agreement by Associated Universities, Inc. This research has made use of the NASA/IPAC Extra-galactic Database (NED), which is operated by the Jet Propulsion Laboratory, California Institute of Technology, under contract with the National Aeronautics and Space Administration. Funding for the SDSS and SDSS-IV has been provided by the Alfred P. Sloan Foundation, the Participating Institutions, the National Science Foundation, the U.S. Department of Energy, the National Aeronautics and Space Administration, the Japanese Monbukagakusho, the Max Planck Society, and the Higher Education Funding Council for England. The SDSS Web Site is \url{http://www.sdss.org/}. This research has made use of the CIRADA cutout service at URL cutouts.cirada.ca, operated by the Canadian Initiative for Radio Astronomy Data Analysis (CIRADA). CIRADA is funded by a grant from the Canada Foundation for Innovation 2017 Innovation Fund (Project 35999), as well as by the Provinces of Ontario, British Columbia, Alberta, Manitoba and Quebec, in collaboration with the National Research Council of Canada, the US National Radio Astronomy Observatory and Australia’s Commonwealth Scientific and Industrial Research Organisation. This publication makes use of data products from the Wide-field Infrared Survey Explorer (WISE), which is a joint project of the University of California, Los Angeles, and the Jet Propulsion Laboratory/California Institute of Technology, funded by the National Aeronautics and Space Administration.

\end{acknowledgements}

% \appendix                  %%appendicial material is supported

% \section{This shows the use of appendix}
% A postscript file is actually an ASCII text file (you may even edit it).
% However, you need to transfer a PDF file or any compressed or packaged
% file in binary mode when using FTP.

% \section{What is SCI?}
% SCI is the abbreviation of Science Citation Index system powered by
% the Institute for Scientific Information (ISI). For details please
% visit {\it http://apps.isiknowledge.com}.

\bibliographystyle{raa}
\bibliography{references}

\end{document}